\documentclass[journal]{vgtc}  

\usepackage{balance}   
\usepackage{xcolor}
\usepackage{colortbl}
\usepackage{amsmath}
\usepackage{longtable}
\usepackage{float} 
\usepackage{bm}
\usepackage{gensymb} 
\usepackage{booktabs}
\usepackage{algorithm}
\usepackage{algpseudocode}
\usepackage{placeins}


\onlineid{0}

\vgtccategory{Research}

\vgtcpapertype{please specify}

\title{Predicting affective connotation of visualizations from their constituent colors}

\author{%
 \authororcid{Karen B.\ Schloss}{0000-0003-4833-4117},
  Halle C. Braun,
  \authororcid{Kushin Mukherjee}{0000-0001-5013-6983},
  \authororcid{Anna L. Chinni}{0000-0003-2269-734X}, and
  \authororcid{Seth R.\ Gorelik}{0000-0002-9186-1767} 
}

\authorfooter{
   \item Karen B. Schloss is with the University of Wisconsin--Madison.
  	E-mail: kschloss@wisc.edu   
    
  \item
  	Halle C. Braun is with the University of Wisconsin--Madison.
  	E-mail: hallebraun43@gmail.com
  \item
  	Kushin Mukherjee is with Stanford University.
  	E-mail: kushinm@cs.stanford.edu

    \item
  	Anna L. Chinni is with the University of Wisconsin--Madison.
  	E-mail: alchinni@wisc.edu

  \item Seth R. Gorelik is with Woodwell Climate Research Center. 
  	E-mail: sgorelik@woodwellclimate.org
}

\abstract{
With increasing evidence that affective connotation (emotional association) is an important aspect of visual communication, there is a need for methods to predict affective connotation of visualizations. Many aspects of visualization design, including colors, textures, and shapes, can contribute to affective connotation, and a key question is how multiple design properties combine to determine the emotion association of a whole visualization. In this study, we focused specifically on color and tested whether it is possible to predict the affective connotation of whole visualizations by aggregating the emotion associations of the individual, constituent colors (additivity hypothesis). We also tested whether accounting for the size of colored regions, as determined by the underlying dataset, improved predictions (data-dependence hypothesis). We found that for colormap data visualizations in which colors were well-distributed across all colors in the color scale, the mean estimated associations of individual colors effectively predicted emotional associations of the maps as a whole (additivity; Exp. 1). For colormaps whose underlying datasets were biased to map more to colors at one end of the color scale, emotional associations were better predicted by a weighted mean that accounted for color frequency in the colormap (data-dependence; Exp. 2). Effects of additivity and data-dependence generalized to dot plots and bar charts (Exp. 3). These results suggest it is viable to predict affective connotation of whole visualizations from their individual design components, which has important implications for automating affective visualization design to support visual communication. }


\teaser{
  \centering 
  \includegraphics[width=1.0\linewidth, alt={}]{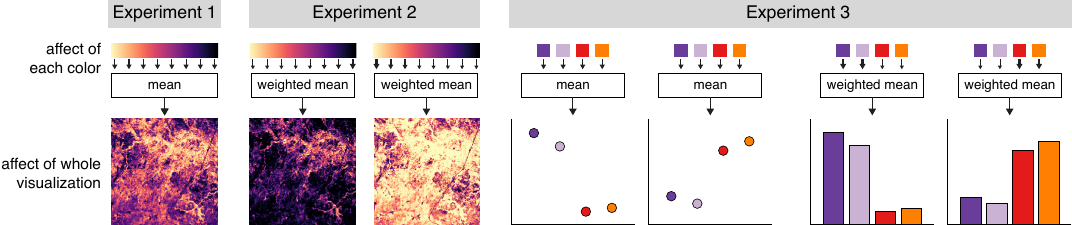}
  \caption{Combining affective connotations of constituent colors to predict affective connotations of whole visualizations, depending on if the relative amount each color appears in the visualization is similar (colormaps in Exp. 1, dot plots in Exp. 3) or differs (colormaps in Exp. 2, bar charts in Exp. 3). When color amount is similar, the simple mean is predictive but when amount differs, a weighted mean accounting for color amount is more predictive, especially when there is greater difference in emotion associations among the colors. }
  \label{fig:teaser}
}

\graphicspath{{figs/}{figures/}{pictures/}{images/}{./}} 

\usepackage{tabu}                      
\usepackage{booktabs}                  
\usepackage{lipsum}                    
\usepackage{mwe}                       

\usepackage{mathptmx}                  

\begin{document}

\firstsection{Introduction} \label{sec:intro}

\maketitle

The affective connotation of visualizations can influence how people evaluate, interpret, and make judgments from visualizations \cite{anderson2022, lan2021, zhang2021}, so there is a growing need for methods to predict affective connotation of visualization designs. Previous work in the domain of color established how variations along color dimensions (i.e., lightness, chroma, yellowness-blueness, redness-greenness) modulate links between color and emotion in visualizations \cite{lin2013, braun2026, blair2025}. For example, color palettes \cite{bartram2017} and color scales \cite{braun2026} that are overall lighter and higher chroma produce visualizations with more positive emotion associations. Moreover, evidence suggests that the amount that each color appears in colormap visualizations, as determined by the underlying dataset, influences emotional associations (data dependence) \cite{braun2026}. For example, colormaps with data values that map predominantly to lighter colors are judged as more positive than colormaps with data values that map predominantly to darker colors, despite being constructed from the same color scale\footnote{The literature varies in terms used to describe (a) color gradations used to represent continuous data (e.g., ``color scales,'' ``color ramps,'' ``colormaps'') and (b) visualizations that map those color gradations onto underlying datasets of continuous data (e.g., ``colormaps'', ``heatmaps''). Following \cite{rogowitz1996, rogowitz2001, schloss2019, sibrel2020, soto2023, schoenlein2023}, we use ``color scale'' to mean the color gradations and ``colormap'' to mean visualizations that map color gradations onto datasets.}. 

\begin{figure}[tb]
 \centering
 \includegraphics[width=0.8\columnwidth]{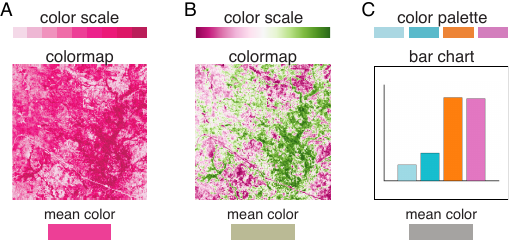}
 \caption{(A) A (roughly) monochromatic color scale, (B) a polychromatic color scale, and (C) a polychromatic color palette with corresponding colormap or bar chart visualizations and mean colors.}
 \label{fig:meanhue}
 \vspace{-6mm}
\end{figure}

Any approach for predicting the affective connotation of a multicolored visualization will need to aggregate over the visualization's colors at some stage of the analysis. In prior work, Braun et al. \cite{braun2026} aggregated at the color appearance level by computing the mean over colors in colormap visualizations along perceptual dimensions in CIELAB/CIELCh color space (lightness (L*), chroma (C*), redness/greenness (a*), and yellowness/blueness (b*)). Next, using these mean values, they predicted variation in affective connotation across colormaps using linear regression models. This approach worked well for the approximately monochromatic (i.e., shades of a single hue) colormaps they studied because the mean color along these dimensions was representative of the colormap as a whole. For example, in Fig. \ref{fig:meanhue}A, the mean color (pink) is representative of the pink/red color scale. 

However, aggregating at the color appearance level can pose major problems for polychromatic (i.e., multi-hue) visualizations, such as the pink/green (PiYG) colormap in Fig. \ref{fig:meanhue}B, in which the mean grayish-green color is not representative of the saturated pinks and greens in the colormap, and the bar chart in Fig. \ref{fig:meanhue}C, in which the mean grayish color is not representative of the cyans, orange, and pink in the visualization. Thus, a new approach is needed to predict affective connotation of visualizations with large hue variability.

In this study, we proposed and tested an alternative approach to predicting affective connotation of whole visualizations that has potential to be robust to wide hue variation. Instead of (1) aggregating over colors at the color appearance level and (2) using the mean color to generate predictions for whole visualizations, our approach (1) estimates emotional associations of each individual color within the visualization and then (2) aggregates over individual color-emotion associations to predict the emotional association for the whole visualization. 

Our approach hinges on the \textit{additivity hypothesis}, by which emotional associations for whole visualizations can be predicted by the aggregate emotional associations of the individual colors. To test this hypothesis, we began by estimating color-emotion associations for each color appearing in a visualization by using regression models of color-emotion associations fitted on a small set of different colors. We used this estimation approach because it was impractical to collect human color-concept association judgments for the large set of colors in the visualizations we studied. We then aggregated these emotion associations over colors within a visualization, using either the simple mean or weighted mean accounting for the prevalence of each color in the visualization. 
Initially, we were skeptical of this approach, assuming that the \textit{whole was different from the sum of its parts}. However, we found this approach was viable for predicting affective connotation of colormap visualizations constructed from a wide variety of color scales (Fig. \ref{fig:scales}; Exps. 1 and 2), as well as dot plots and bar charts constructed from a variety of color palettes (Fig. \ref{fig:palettes}; Exp. 3). 

\textbf{Contributions.} This paper makes the following contributions: (1) We present an effective approach for predicting color-emotion associations for whole visualizations from estimated associations of their constituent colors, (2) We show that our approach works well using associations estimated from pre-fit regression models without needing to collect new association data for individual colors, (3) We show that our approach generalizes across visualization types---colormap visualizations for continuous data, dot plots and bar charts for categorical data, and (4) We provide empirical evidence supporting the additivity and data-dependence hypotheses for understanding affective connotation of visualizations. In the General Discussion, we discuss how our method can be deployed to estimate affective connotation based on color for visualizations more broadly, and consider limitations to be addressed before building a full-fledged automated tool.

\section{Background}

\subsection{Links between color and emotion for visualization} 

Extensive research has studied links between colors and emotions \cite{jonauskaite2025, schloss2020, wright1962, adams1973, dandrade1974, dael2016, valdez1994}, with a recent rise in such studies specifically for visualization \cite{bartram2017, blair2025, anderson2022, braun2026}. These studies have focused on either color palettes for visualizations of categorical data  \cite{bartram2017, anderson2022, blair2025} or color scales for visualizations of continuous data \cite{zhang2021, samsel2018ArtAffect, braun2026}, reviewed below. 

\textbf{Color palettes for visualizations of categorical data.}
In their seminal work on affective connotation of color palettes, Bartram et al. \cite{bartram2017} investigated associations between palettes and emotions that span the pleasure (valence) and arousal (activation) axes of the pleasure, arousal, dominance (PAD) model of affect \cite{russell1980,posner2005}. They studied the correspondence of emotion terms to color dimensions (lightness, chroma, and hue) and found effects largely for lightness and chroma. The terms \textit{positive}, \textit{playful}, and \textit{calm} were associated with lighter colors, \textit{disturbing}, \textit{serious}, and \textit{negative} were associated with darker colors,  \textit{exciting}, \textit{playful}, and \textit{positive} were associated with higher chroma colors, and  \textit{calm} was associated with lower chroma colors. They also found effects of hue, such as \textit{calm} associated with cooler hues (blues and greens) and \textit{exciting} associated with warmer hues (reds, oranges, yellows). Blair et al. \cite{blair2025} found similar effects when they investigated self-reported affective state in response to Bartram et al.'s color palettes.  

 Blair et al. \cite{blair2025} also compared affective state in response to color palettes displayed as abstract squares vs. points in scatter plots and found similar results. They interpreted this similarity as evidence that color palettes studied outside the context of visualizations provide meaningful results for visualization, implying emotionality of colors is agnostic to data display without need for data-awareness. But, the colors within color palettes had similar affect (e.g., all relatively positive or negative) and appeared in similar proportions within the visualizations. As discussed in the next section and in our study, data-awareness becomes more relevant when colors vary in affect and proportion \cite{braun2026}.

Extending Bartram et al. \cite{bartram2017}, Anderson and Robinson \cite{anderson2022} investigated downstream effects of affective connotation on interpretations and evaluations of visualizations. They tested for effects of affective congruence, in which visualizations paired color palettes with data topics of similar emotional association (e.g., positively associated light, saturated colors with positive topics like types of dessert as opposed to negative topics like types of homicide). Affectively congruent visualizations were judged as more appropriate, and congruence amplified the emotionality of visualizations (e.g., maps of positive topics were judged as especially pleasurable when designed using positive colors).  

\textbf{Color scales for colormap visualizations of continuous data.}
In colormap visualizations, a gradation of color---a color scale---is mapped on to a gradation of magnitude, typically to illustrate spatial variation in a dataset\cite{brewer1994, hegarty2011}. In initial work aiming to link colormaps to emotions, Samsel et al. \cite{samsel2018ArtAffect} designed color scales inspired from colors in paintings under the premise that colormaps would share the affective connotation of paintings from which the colors were sourced. 

Subsequently, Braun et al. \cite{braun2026} investigated whether it is possible for colormaps to convey strong affective connotation while also having characteristics needed to reveal important patterns in the data. Their logic was that affective connotation of colors is largely driven by overall lightness (e.g., lighter colors are more associated with positive emotions) \cite{wright1962, dandrade1974, schloss2020, jonauskaite2025,  bartram2017}, yet \textit{variation} in lightness is important for aspects of spatial vision (e.g., contrast sensitivity in high spatial frequencies, orientation sensitivity, pattern and shape recognition) \cite{devalois1990, devalois1993, rogowitz1996, kindlmann2002, livingstone1987} needed to detect detailed spatial structure like fine details in geographical maps \cite{ware1988, rogowitz1996}. Thus, Braun et al. \cite{braun2026} tested whether colormaps that have strong lightness contrast to support spatial vision can also convey systematic affective connotation. Participants rated emotional associations for colormaps that varied in lightness and were (approximately) monochromatic, differing in their overall hue (red/yellow/green/blue), overall lightness (light/dark), and overall chroma (high/low). They found that colormaps with strong lightness contrast indeed had systematic emotion associations, primarily driven by overall lightness, but also influenced by chroma, yellowness/blueness, and redness/greenness. Analysis of emotion dimensions (valence: positive/negative, and activity: activated/deactivated) showed that higher valence corresponded to lighter, higher chroma, and bluer colormaps, and higher activation corresponded to lighter, yellower, and redder colormaps.

Returning to the issue of data awareness, Braun et al. \cite{braun2026} also investigated whether affective connotation of colormaps depended solely on \textit{which} colors were in the color scale used to construct the colormaps (data agnostic hypothesis) or depended also on \textit{how much} those colors appeared in the colormaps as determined by the underlying dataset (data-dependence hypothesis). To adjudicate these hypotheses, they had participants rate emotional associations of colormaps constructed from the same color scales, applied to datasets that were ``shifted'' to have more data values map to the darker (more negatively associated) endpoint of the color scale (``dark shifted'') or the lighter (more positively associated) endpoint (``light shifted''). Supporting the data-dependence hypothesis, shifting the underlying datasets influenced the emotional associations of colormaps produced from the same color scales (e.g., light shifted colormaps were associated more with \textit{positive} and less with \textit{negative} or \textit{calm} than dark shifted maps). In some cases, shifting the dataset moved colormaps from having positive valence to negative valence. Overall, color-emotion associations were better predicted by data-aware models that weighted colors in color scales by how much they appeared in colormaps, compared to data-agnostic models that gave equal weight to each color \cite{braun2026}. These results challenge the generality of Blair et al.'s \cite{blair2025} suggestion that links between color and emotion for visualization can be studied outside of the context of visualizations, and are consistent with broader arguments for the need for data-aware design (see \cite{zeng2021}).

\subsection{Additivity of judgments about color} \label{sec:additivity}
As introduced in Section 1,  our approach hinges on the additivity hypothesis, by which the emotional associations of individual colors in a visualization add up to produce the emotional associations of the whole visualization. Thus, in this section, we review prior literature on when evaluative judgments about color are/are not additive. 

One case in which judgments about color are \textit{not} simply additive is predicting aesthetic preference for color pairs based on preference for the individual colors \cite{schloss2011, ou2004_pairs}. For example, Schloss and Palmer \cite{schloss2011} found that only 21.7\% of the variance in average preference for 992 color pairs was predicted by preference for the individual colors, but the total variance explained increased to 62.9\% when accounting for the overall coolness of the colors (cooler preferred), hue distance (more similar hues preferred), and lightness distance (greater lightness contrast preferred). Critically, the latter two factors are \textit{relational} variables that compare colors within the combination and cannot be computed for colors in isolation.  

Although preferences for individual colors do not strongly predict preferences for color pairs, preferences for color pairs do predict preference for $>2$ color combinations \cite{gramazio2017}, suggesting additivity of higher-order color combinations. That is, accounting for relational factors in color pairs (i.e., hue similarity, lightness contrast), is effective for predicting preference for color combinations comprised by those pairs. Indeed, when developing Colorgorical as a tool to automate palette generation, Gramazio et al. \cite{gramazio2017} found that participants' aesthetic preferences for visualizations of 3, 5, or 8 colors were well-predicted by model estimated preferences computed using the regression equation for color pairs in Schloss and Palmer \cite{schloss2011}. Likewise, Ou et al. \cite{ou2011} found that judgments of color harmony for shapes resembling a pie chart were well-predicted by pairwise estimates of color harmony. 

However, associations between colors and concepts, like emotions, may be more additive than  color preferences \cite{ou2004_pairs}.
 Ou et al. \cite{ou2004_pairs} found that the mean associations for individual colors predicted associations for color pairs when evaluated for the following dimensions: warm-cool, heavy-light, modern-classical, dirty-clean, active-passive, hard-soft, tense-relaxed, fresh-stale, and masculine-feminine (referred to as ``color emotions''). These findings support the possibility that color-emotion associations for typical emotions like angry, fearful, disgust, happy, and sad may also be additive, such that emotional associations of whole visualizations are well-predicted by the additive combination of their constituent colors. We tested this possibility in our study.

\subsection{Using color space regression to describe and predict patterns of data over colors} \label{sec:color_space_regression}
To test whether we could predict color-emotion associations for whole visualizations from their constituent colors, it was necessary to quantify associations for the constituent colors. In our study, we estimated these associations using \textit{color space regression}, so we review the literature on color space regression here. 

Color space regression emerged as a solution to a general problem in studies of color cognition, which is to describe and predict patterns of data over a variety of colors. Much of the work addressing this challenge has been done in the study of color preferences, aiming to describe patterns of color preferences among groups of participants or individuals \cite{hurlbert2007, ling2007, ou2004_single, palmer2010, bimler2014, schloss2015, schloss2017, schloss2018modeling}. Color space regression is a powerful approach because it uses existing color coordinates in a specified color space as predictors in a regression model fit to a given dataset. The resulting regression equation can be used both to \textit{describe} the pattern of data to which it was fit and to \textit{predict} data for colors not included in the original regression by using the color coordinates as input (so long as the regression equation is a good fit to the data).

Initial work successfully used color coordinates in cone-contrast space to predict color preferences \cite{hurlbert2007,ling2007}, but a systematic study of a variety of color spaces and ways of specifying coordinates within spaces found that a model in CIELAB/CIELCh space was most predictive of color preferences \cite{schloss2018modeling}. The most performative model was one that included parameters for lightness (L*), chroma (C*), in addition to two pairs of parameters that captured variation in hue, h. The first pair was sine(h) and cosine(h) (first harmonic), which together estimate a dominant hue. This first harmonic captures a peak at a given hue angle and trough at the opposite hue angle, such as the peak around blue and trough around yellow characteristic of mean color preference data. The second pair was the sine(2h) and cosine(2h) (second harmonic), which together estimate a dominant hue axis. This second harmonic captures peaks among opposite hues (e.g., reds \textit{and} greens) and troughs among hues rotated 90 degrees (e.g., blues \textit{and} yellows). This model was referred to as the ``LabC Cyl2'' model because it uses cylindrical coordinates in CIELAB space and includes two harmonics. 

The LabC Cyl2 model is not only effective for describing and predicting color preferences \cite{schloss2018modeling}, but also for describing patterns of color-concept associations used to study how people interpret color meaning in data visualizations (Supplementary Material of \cite{schoenlein2023}). Most recently, Mukherjee et al.\cite{mukherjee2022color} used this model to describe the underlying dimensions of color semantic space for a wide variety of concepts, including the emotions: angry, fearful, disgust, happy, and sad. They also found that LabC Cyl2 captures most of the variance in color-concept association ratings for these emotions  angry $R^2 = 0.878$, disgust $R^2 = 0.852$, fearful $R^2 = 0.737$, happy $R^2 = 0.837$, and sad $R^2 = 0.828$. We used these models to estimate color-emotion associations for each color in the color scales (Fig. \ref{fig:scales}) and color palettes (Fig. \ref{fig:palettes}) in the present work.


\section{Experiment 1}
\begin{figure}
 \centering
 \includegraphics[width=0.9\columnwidth]{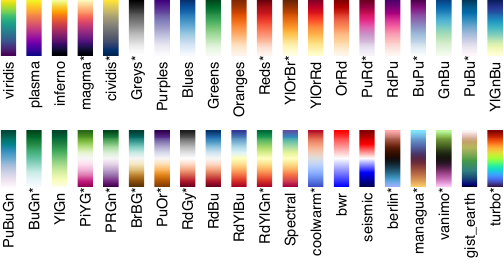}
 \caption{40 color scales from the Matplotlib Python library \cite{Hunter2007} used to construct the colormap visualizations in Exp. 1 (*20 scales in Exp. 2.)}
 \label{fig:scales}
 \vspace{-6mm}
\end{figure}

Exp. 1 tested whether it is possible to predict the affective connotation of whole colormap data visualizations from the simple mean affective connotation of individual colors (additivity hypothesis). The colormaps were constructed from underlying datasets that were well-distributed over the full range of values, such that all colors within the color scale appeared in similar proportions within the visualizations (from \cite{braun2026}). We sought to test colormaps designed using a variety of color scales commonly used for visualization, so we sampled 40 representative color scales from the \texttt{Matplotlib} plotting library \cite{Hunter2007} (Fig. \ref{fig:scales}). With 40 color scales and 256 colors per color scale (10,240 colors total), it was impractical to collect human ratings on the affective color-emotion associations for each individual color, so we estimated associations using color space regression (Section \ref{sec:color_space_regression}).
The experimental stimuli, data, and analysis scripts for all experiments in this study can be found in our GitHub repository\footnote{\url{github.com/SchlossVRL/color_scales_affect_predict}}.

\subsection{Methods}

\textbf{\indent Participants.}
Our target sample was $n=80$ to balance the assignment of color scales to underlying datasets across participants (see Design and displays). Due to the nature of condition assignment during simultaneous online data collection, we collected a total of 107 participant datasets, but only analyzed the data from the first 80 participants needed to balance the conditions and who had typical color vision. The participants in this and all subsequent experiments were undergraduates from the University of Wisconsin--Madison who gave informed consent and received extra credit in their Psychology course for their participation (protocol approved by the UW--Madison IRB). They provided demographic information in free-response text boxes. We assessed color vision using (1) two self-report questions asking if they had difficulty distinguishing between colors relative to the average person and if they considered themselves colorblind (excluded if they reported `yes' to at least one question) and (2) using digital Ishihara plates (excluded if incorrectly identified the number in more than 1 out 11 Ishihara plates).
In this experiment, the mean age was 19.14 (range:18 -- 22) and their genders included 76 women, 29  men, 1 non-binary, and 1 other. Their race/ethnicity included 75 White, 8 Asian, 2 African American, 4 Arab, 6 Latino/Hispanic, 1 Malian American, 2 Indian, 1 Middle Eastern, 1 Vietnamese, 1 Greek, 1 Black, 1 White/Black, 2 Hispanic/White, 1 White/Middle Eastern, 1 did not report. Five participants were excluded for atypical color vision.

\textbf{Design and displays.} 
\label{sec:exp1_design_displays}
During the experiment, all participants judged 80 colormap data visualizations for each of 5 emotions: angry, disgust, fearful, happy, and sad. We chose these five emotions out of all possible emotions because there already existed a color-emotion association dataset for these emotions and colors broadly sampled over color space \cite{mukherjee_llm}, which we could use to generate the model predictions. 

We constructed the 80 colormaps by applying each of the 40 color scales in Fig. \ref{fig:scales} to 80 different underlying datasets, once when the upper end of the scales in Fig. \ref{fig:scales} was mapped to larger values in the dataset (Matplotlib default) and once in the reversed mapping to ensure the results would not be due to the specific mapping. The underlying datasets represent global raster data of aboveground live woody biomass (AGB) \cite{harris2021}, and are a subset of those used to generate colormaps tested in Braun et al. (2026) \cite{braun2026} (see Supplementary Material Fig. \ref{fig:graymaps44} and Section \ref{sec:underlyingdata}). Starting with the 10 underlying datasets that had the most similar affective connotation when presented as gray scale colormaps \cite{braun2026}, we created eight variants per map by rotating them clockwise (0\degree, 90\degree, 180\degree, 270\degree) and then mirror reflecting them. 

Each participant saw a unique pairing of 40 color scales in default and reversed mapping (80 color scale conditions) to each of 80 underlying datasets, so no participant saw colormaps constructed from the same underlying dataset in more than one color scale. We achieved this unique pairing using a Latin square design, which pre-generated the stimuli for each participant by assigning each of the 80 color scale conditions to one of the 80 underlying datasets. 80 participants were needed to pair each color scale condition with each underlying dataset.  

We presented the trials in a block randomized design, such that all participants judged all 80 colormaps for each emotion in a random order before going on to the next emotion (emotions were also randomized). The randomization was applied dynamically for each participant. Each experiment trial displayed one colormap centered on the screen scaled to be 200px × 200px. This size corresponded to 3.8cm × 3.8cm on a display that was 29.5cm × 16.5cm with a resolution of 1920px × 1080px. Below was a slider scale with the endpoints labeled `not at all' and `very much' and an unlabeled midpoint. The background of the monitor was gray (rgb: [128,128,128]). All experiments were presented using jsPsych \cite{de2015jspsych, de2023jspsych} and participants accessed the experiment through browsers on their personal computers. Thus, the exact colors each participant saw depended on properties of their own monitors. All reported CIELCh values are approximations assuming a sRGB display (typical in visualization research that aims to produce results that are robust to display variation \cite{stone2014, szafir2018, gramazio2017, mukherjee2022}).

\begin{figure}[tb]
 \centering
 \includegraphics[width=1.0\columnwidth]{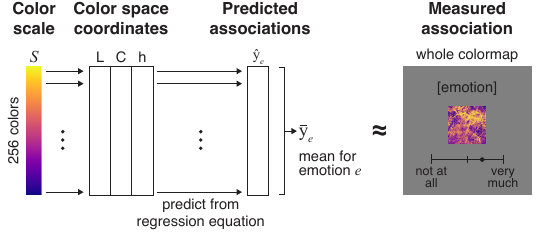}
 \caption{Process to estimate emotion associations for each color within a color scale and test if the mean estimated association predicts human emotion associations for whole colormaps (see text for details).}
 \label{fig:process}
 \vspace{-6mm}
\end{figure}

\textbf{Procedure.}
The participants were first told that they would be presented with a series of colormaps, one at a time, and their task was to rate how much they associated each map with each of the five emotion concepts on a scale from `not at all' to `very much' (Fig. \ref{fig:process}, right). So they could anchor what associating `not at all' and `very much' meant to them in the context of this experiment \cite{palmer2013visual}, they were shown a list of all five concepts with colormaps from each of the 40 color scales, and were asked to consider which colormap they associated most/least with each emotion concept. They were asked to rate those colormaps near the `very much' and `not at all' endpoints of the response scale (respectively) during the experiment, and to use the full range of the scale. They were also informed they would be asked to rate each map for a given concept before starting the next concept, and were asked to make ratings based on their first intuition because we were interested in their initial impressions. On the next screen, the participants were instructed to use the response scale by sliding the cursor to the position they wanted to select and letting it go to record their response. They completed four training trials instructing them to move the slider to each endpoint and halfway between the center and each endpoint. If they did not place the slider near the instructed region, they repeated the training trial until the correct region was selected. The participants then completed five blocks of trials, one for each emotion. Before each block they were told which emotion they would be asked to judge and then they were presented with the 80 trials within that block, each separated by a 250ms inter-trial interval.

\begin{figure*}[tb]
 \centering
 \includegraphics[width=.9\textwidth]{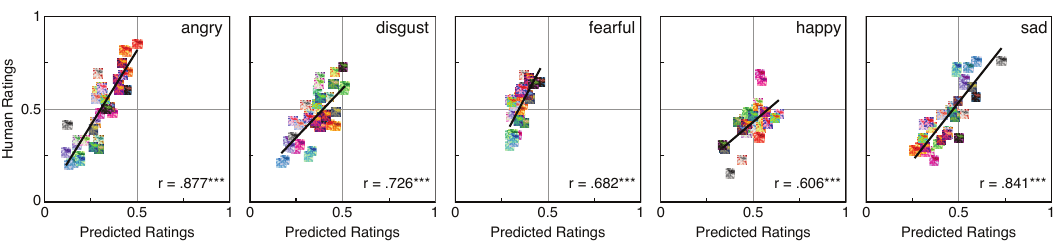}
 \caption{Scatterplots showing the mean human color-emotion association ratings for colormaps constructed from each of the 40 color scales as a function of the predicted associations based on the mean estimated associations for the individual colors. Subplots show the results for each emotion, with the points showing example colormaps for each color color scale. ***$p<.001$.}
 \label{fig:exp1_results}
 \vspace{-6mm}
\end{figure*}

\subsection{Results and discussion}

To test whether the average affective connotation of individual colors in each color scale predicted affective connotation of corresponding colormaps as wholes, we first needed to generate estimates of the affective connotation of the individual colors. This process is illustrated in Figure \ref{fig:process}. For each of the 40 color scales $S$, we converted each of their 256 colors in RGB space to CIELCh coordinates. Then, for each of the five emotions $e$, we estimated the color-emotion associations $\hat{y}$ for each individual color using Eq. \ref{eq:color_space_reg} with the coordinates in CIELCh space and the values of $w_1 - w_7$ in Supplementary Material Table \ref{tab:reg_weights}. 

\vspace{-1.5em}
\begin{equation}
\hat{y} = w_1 + w_2L + w_3C + w_4\sin h + w_5\cos h + w_6\sin 2h + w_7\cos 2h
\label{eq:color_space_reg}
\end{equation}
\vspace{-1.5em}

The weights in Table \ref{tab:reg_weights} were obtained by applying the LabC Cyl2 color space regression model (Section \ref{sec:color_space_regression}) to human color-concept association rating data in Mukherjee et al. \cite{mukherjee2022}. These data were collected by having participants rate the association between each color from a set of 71 colors called the ``UW-71,'' systematically sampled over  CIELAB color space  (see Supplementary Material Section \ref{sec:UW71} and Fig. \ref{fig:UW_71}), and each of the five emotion concepts. 

We next asked whether these regression equations, fit on just 71 colors, can estimate color-emotion associations for the 256 colors in each of the 40 color scales (Fig. \ref{fig:scales}) well enough to predict human color-emotion association ratings for whole colormaps produced by those scales. To address this question, we first aggregated the predicted associations across all colors within each color scale for each emotion $\bar{y}_e$ by computing the mean. We then tested whether these predictions correlated with the mean color-concept associations for colormaps produced from those color scales, averaged over participants (Fig. \ref{fig:process}) using Pearson's \textit{r} in these (and all subsequent) correlation analyses. Fig. \ref{fig:exp1_results} shows that the predicted associations were significantly correlated with human ratings for all five emotions: angry $r = 0.877$, disgust $r = 0.726$, fearful $r = 0.682$, happy $r = 0.606$, and sad $r= 0.841$ (all df = 38, all $p$s $< .001$) (Supplementary Fig. \ref{fig:Exp 1 Histograms} shows histograms of correlations at the participant level and Fig. \ref{fig:dataset_cor_histograms} shows the mean ratings). However, the strength of the correlations varied across emotions, which may be due to some emotions (e.g., fearful and happy) having a more restricted range in human ratings compared to other emotions (e.g., angry and sad), resulting in less variability for the predictions to capture. 

We also note that while the correlations show that the predicted ratings matched the pattern of human associations, the predicted ratings generally underestimated the strength of emotional associations. This scaling issue could be mitigated by using regression to fit the slope and intercept to the human ratings, which would not affect the overall model fit but would help predict more accurate association magnitudes if absolute rather than relative predictions were needed.

\textbf{Summary.} Supporting the additivity hypothesis, Exp. 1 showed it is possible to predict emotional associations of whole colormap visualizations from the mean emotional association of the constituent colors. Moreover, it was possible to do so using estimated associations from previously fit regression equations, suggesting our approach has potential to scale to predict affective connotation of any colormap visualization without needing to collect additional data from human participants. However, the datasets used in Exp. 1 were generated such that data values were distributed approximately uniformly. Thus, colors along the full range of the color scales appeared approximately equally within the map. Colormaps that have disproportionate amounts of colors might require aggregation methods that go beyond simple averaging of estimated ratings. We tested this possibility in Exp. 2. 

\section{Experiment 2}
The data-dependence hypothesis predicts that when visualizations contain disproportionate amounts of some colors relative to others, the colors that occupy more space will have a greater influence on the affective connotation \cite{braun2026}. We tested this hypothesis by having participants rate emotion associations of colormaps that had shifted underlying datasets (i.e., skewed to map more to colors toward one endpoint or the other within each color scale). We evaluated whether these association ratings were better predicted by the mean association of all colors in the map equally weighted (as in Exp. 1) or if they were better predicted by a weighted mean that accounted for how much each color appeared in the visualization (data-dependence). 

\begin{figure*}[tb]
 \centering
 \includegraphics[width=.9\textwidth]{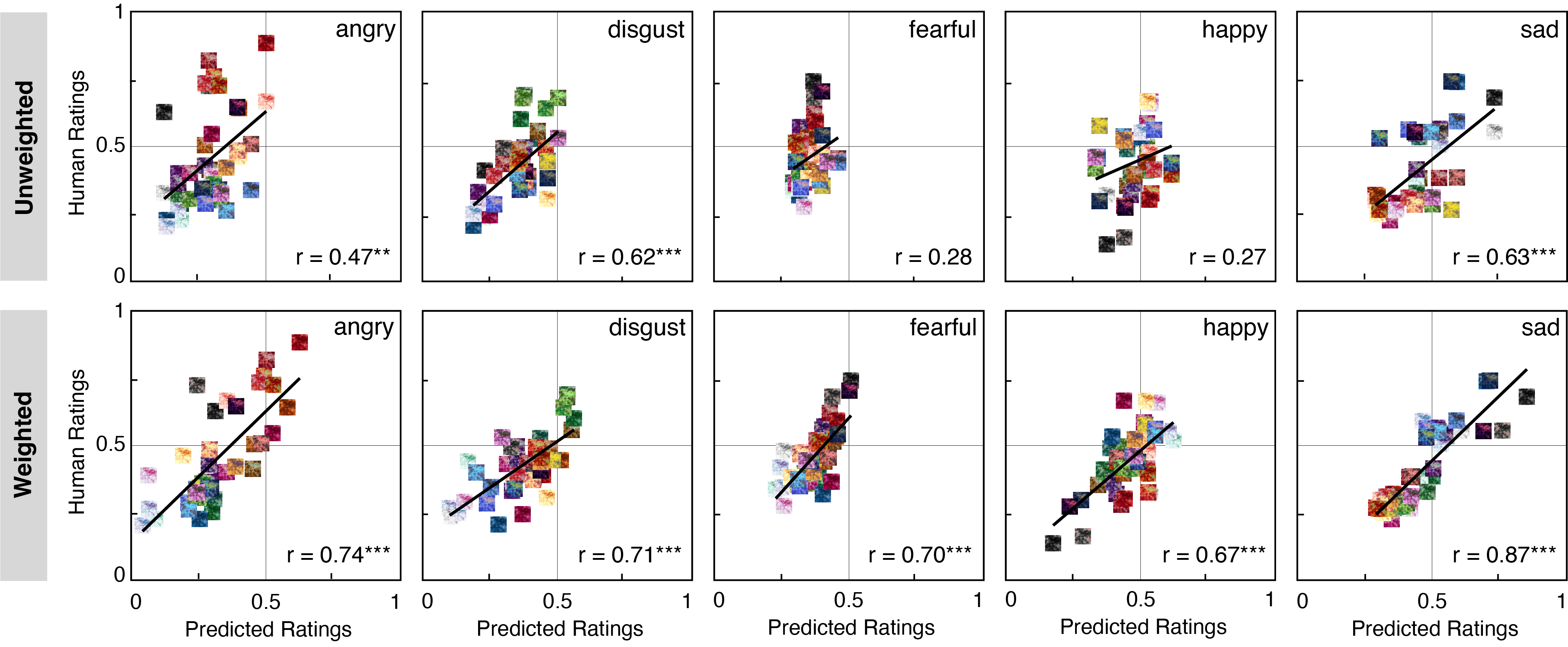}
 \caption{Scatterplots showing the mean human color-emotion associations for shifted colormaps as a function of the unweighted mean color-emotion associations of the constituent colors within each color scale (top row) or associations weighted by pixel count (bottom row).  **$p < .01$ ***$p < .001$.}
 \label{fig:Exp2 Results}
 \vspace{-5mm}
\end{figure*}

\subsection{Methods}

\textbf{\indent Participants.}
We recruited $n=40$ to balance the assignment of color scales to underlying datasets for each colormap. Mean age was 19.18 (range: 18 -- 22). Self-reported genders included 34 women, 5 men, and 1 other. Their race/ethnicity included 26 White, 2 Hispanic, 6 Asian, 2 Black, 1 South Asian, 1 North African, 1 Latino/Hispanic, and 1 Indian. All had typical color vision.


\textbf{Design, displays, and procedure.}
The design, displays, and procedure were the same as Exp. 1 except for the following adjustments. First, we tested shifted versions of the underlying datasets from Exp. 1 (from Braun et al. \cite{braun2026}; see Supplementary Material Section \ref{sec:supp-shift} of the present paper for how shifts were computed). In Braun et al., all color scales varied monotonically in lightness, so they referred to the shifted colormaps as ``dark''/``light'' shifted when underlying datasets were skewed to produce darker/lighter colormaps, respectively. Here, we included divergent color scales (not varying monotonically in lightness), so the dark/light shift terms did not pertain. Instead, we refer to the skewed datasets as ``high''/``low'' shifted (i.e., shifted toward high or low values in the dataset). Second, we reduced the number of color scales from 40 to 20 by eliminating those that were most similar to one another in their emotional associations across all five emotions (see Supplementary Material Section \ref{sec:sub-sampling}). This change enabled accommodating the additional variable of shift without increasing the number of trials from Exp. 1. Third, we controlled for variability in the underlying datasets by assigning the same color scale to the high and low shifted version of datasets for each participant. As such, only 40 participants were needed to fully balance assignments of color scales to underlying datasets. The combination of 20 color scales $\times$ 2 shifts (left/right) $\times$ 2 mappings (default/reversed) produced 80 colormaps, which participants judged for each of the 5 emotion concepts (400 trials).

\subsection{Results and discussion}
To analyze the data, we generated two sets of predictions for the colormaps. The unweighted predictions were simply the mean of the estimated color-emotion associations as in Exp. 1. We computed the weighted predictions by multiplying the estimated color-emotion association for each color by the proportion of pixels in which that color appeared in the colormap before computing the mean. Fig. \ref{fig:Exp2 Results} shows the mean emotion associations for high and low shifted colormaps generated for each color scale\footnote{We averaged over mapping and shift conditions that were analogous: default mapping applied to high-shifted datasets with reversed mapping applied to low-shifted datasets, and reversed mapping applied to high-shifted datasets with default mapping applied to low-shifted datasets.} as a function of the unweighted predictions (Fig. \ref{fig:Exp2 Results}, top row) or weighted predictions (Fig. \ref{fig:Exp2 Results}, bottom row). 

Supporting the data-dependence hypothesis, the mean human ratings correlated more strongly with the weighted predictions than the unweighted predictions numerically for all five emotions (\textit{r} values shown in Fig. \ref{fig:Exp2 Results}). The weighted predictions were significantly more predictive for angry ($t = 2.29, p =.014$), fearful ($t = 2.73, p = .005$), happy ($t = 2.86, p = .004$), and sad ($t = 3.53, p = .001$).47, but the difference was not significant for disgust ($t = .85, p = .200$) (1-tailed, given the data dependence hypothesis predicted directionality). Histograms of correlations at the participant level are in Supplementary Fig. \ref{fig:Exp 2 Histograms}.

We further analyzed the data at the participant-level using two mixed-effect linear regression models for each emotion, one that used the unweighted predictions and one with the weighted predictions (both centered).  The models predicted participant ratings for each trial for a given emotion, including by-participant random intercepts and slopes to capture individual-level variance in ratings (see Supplementary Material Table \ref{tab:supp-exp2-models} for full model results). 
To compare model fits, we used the Akaike Information Criterion (AIC) \cite{akaike1974new} and the Bayesian Information Criterion (BIC) \cite{schwarz1978estimating} (lower values indicate better fit). AIC is based on the maximized log-likelihood for the model given the data, penalizing fit for each additional parameter. BIC is similar, with a slightly stricter penalty for additional parameters that increases with sample size. Supporting the data-dependence hypothesis, AIC and BIC were lower (better fit) for the weighted prediction model than the unweighted prediction model for all five emotions (Table \ref{tab:exp2-uniform-v-weighted}).

\begin{table}[t!]
\centering
\small	
\caption{Comparing unweighted models to weighted models. The lower AIC/BIC values for each model are indicated using $\downarrow$ symbols. Weighted models consistently show better fits to participant ratings.}
\label{tab:model_comparison}
\begin{tabular}{lllll}
\toprule
 & \multicolumn{2}{c}{Unweighted} & \multicolumn{2}{c}{Weighted} \\
\cmidrule(lr){2-3} \cmidrule(lr){4-5}
Emotion & AIC & BIC & AIC & BIC \\
\midrule
Angry & 736.1 & 772.5 & 170.1 $\downarrow$& 206.5 $\downarrow$\\
Disgust & 676.6 & 713.0 & 566.0 $\downarrow$ & 602.4 $\downarrow$\\
Fearful & 782.4 & 818.8 & 498.0 $\downarrow$& 534.4 $\downarrow$\\
Happy & 563.3 & 599.7 & 227.1 $\downarrow$& 263.5 $\downarrow$\\
Sad & 201.1 & 237.5 & -382.5 $\downarrow$& -346.1 $\downarrow$\\
\bottomrule
\label{tab:exp2-uniform-v-weighted}
\vspace{-3.5em}
\end{tabular}
\end{table}

\textbf{Summary.} Exp. 2 showed that affective connotation of colormaps with disproportionate amounts of each color can be predicted by affective connotation of the individual colors weighted by their prevalence in the visualization. These findings are consistent with the data-dependence hypothesis, and emphasize the importance of data-aware visualization design. However, there are limitations of modeling data-aware design at the pixel level, as discussed in the General Discussion.

\section{Experiment 3}
In Exp. 3 we investigated whether our approach established for colormap visualizations of continuous data constructed from sequential or diverging color scales (Exps. 1 and 2) extends to visualizations of categorical data constructed from qualitative color palettes (Fig. \ref{fig:palettes}). We tested 4-color palettes applied to dot plots and bar charts representing the same underlying datasets (Fig. \ref{fig:teaser}), with two data points having high y-values and two having low y-values. Thus, for both dot plots and bar charts, the 4 data marks had different positions, but the size of the data marks were equal in dot plots and varied in bar charts. The additivity hypothesis implies that emotion associations for both dot plots and bar charts will be predicted by the mean emotion associations of the four colors within the palettes. We considered two forms of the data-dependence hypothesis. If the size of colored regions is the factor that drives data dependence (as suggested in Exp. 2 and \cite{braun2026}), then the y-coordinate of the data marks will influence affective connotation for bar charts but not dot plots because in bar charts (but not dot plots), the y-coordinate influences color region size. However, if mark position also contributes to data-dependence when judging affective connotation (e.g., greater influence for higher data marks), then weighting by y-position could also augment predictions for dot plots.

\subsection{Methods}

\textbf{\indent Participants.}
The participants were assigned to the dot plot ($n=40$) or bar chart ($n=41$) condition. Overall, their mean age was 19.16 (range = 18-22). Their reported gender included 72 women, 7 men, and 2 non-binary. Their race/ethnicity included 38 White, 8 Asian, 6 Black, 5 Hispanic, 2 Asian American, 2 Indian, 2 Latino/Latina, 1 African American, 1 South Asian, 1 Colombian, 1 Middle Eastern, 1 Mixed, 1 Malian American, and 1 Chinese (some participants indicated $>1$ race/ethnicity). We excluded 1 participant (bar chart condition) for atypical color vision.

\begin{figure}[tb]
 \centering
 \includegraphics[width=1.0\columnwidth]{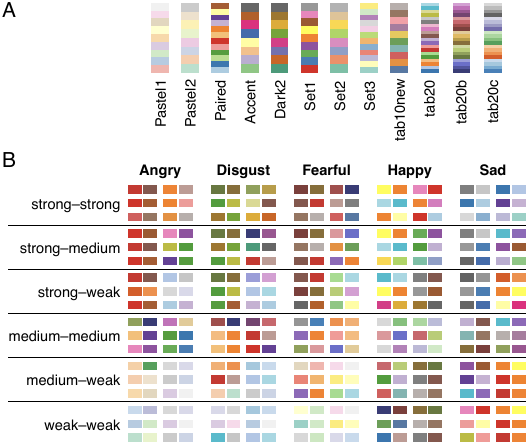}
 \caption{(A) Qualitative color palettes from Matplotlib including ColorBrewer and Tableau palettes. (B) 4-color palettes in the present study tested for each emotion, comprised by the combinations of strong, medium or weakly associated colors for each emotion (see text for details).  }
 \label{fig:palettes}
 \vspace{-6mm}
\end{figure}

\textbf{Design and displays.}
The stimuli were bar charts and dot plots produced from 18 4-color palettes for each of five emotions (angry, disgust, fearful, happy, and sad) (Fig. \ref{fig:palettes}B). Each palette contained two pairs of colors, with the pairs having strong, medium, or weak associations with the given emotion. All pairwise combinations of these three emotion strengths produced six strength conditions (strong--strong, strong--medium, strong--weak, medium--medium, medium--weak, and weak--weak). We tested three versions of each strength condition, resulting in the 18 palettes in Fig. \ref{fig:palettes}B. For example, the first three rows of Fig. \ref{fig:palettes}B include colors that are all strongly associated with each emotion, and the second three rows include one pair of colors that is strongly associated (left pair) and the other pair that is medium associated (right pair) with each emotion. 

To select the 4-color palettes to achieve this experimental design, we began with the qualitative color palettes in Matplotlib \cite{Hunter2007}, which included palettes from ColorBrewer and Tableau (replacing the original Tableau 10 with the new Tableau 10). Next, we used the regression equation in Eq. \ref{eq:color_space_reg} and weights in Table \ref{tab:reg_weights} to estimate associations between each color within each palette and each emotion. We then used the estimated associations to select color palettes for each emotion using the following procedure. First, we defined target association strength vectors for each of the six association strength conditions, coding ``strong'' as 1, ``medium'' as .5, and ``low'' as 0 (e.g., the vector for strong-medium was coded as [1 1 .5 .5]). Second, for each of the 12 palettes in Fig. \ref{fig:palettes}A, we determined the four colors whose association strength had the minimum mean squared error (MSE) for each target association vector. Third, for each of the 6 association strength conditions, we chose three 4-color palettes that had the lowest MSE from the target association vector for that condition.\footnote{To avoid palettes appearing like sequential color scales, we skipped palettes that contained $>2$ colors of the same general hue and went to the palette with the next lowest MSE. This rule applied to 5/90 palettes (always tab20b or tab20c).}     

During the experiment, participants judged four visualizations for each color palette determined by two factors: association order (descending/ascending) and y-position (left-high, right-high). For descending association order, the colors were ordered along the x-axis from most (left) to least (right) strongly associated with the emotion for that block (ascending order was the opposite). When y-position was left-high, the left two data points had greater y-values than the right two data points (right-high was the opposite). For example, in Fig. \ref{fig:teaser}, the data points are descending in association strength for the emotion sad, and they are left-high in the left plot and right-high in the right plot. The full experimental design included 6 emotion strength parings $\times$ 3 versions of each pairing $\times$ 2 association orders (ascending/descending) $\times$ 2 y-value positions (left-high/right-high) $\times$ 5 emotions, producing 360 trials. The participants judged all 72 visualizations in their assigned plot type (dot/bar) for each emotion in a random order before going on to the next emotion. Emotion order was also randomized.  

To generate the underlying datasets of four values for each plot, we defined the ``high'' y-values as 80 and the ``low'' y-values as 20 on a scale from 0 to 100 and added a random jitter of +/- 10 pixels to each of the four values. Using this procedure, we generated unique datasets for each plot for each participant within a given visualization condition, but underlying datasets were the same for the dot and bar conditions (i.e., subject 1 in the dot condition saw the same underlying datasets mapped to the same colors as subject 1 in the bar condition). 

The dot plots and bar charts were 7 cm wide x 5.8 cm tall, centered on a 9.1 cm x 8 cm white background when displayed on a 29.8 x 16.5 cm monitor (1920 $\times$ 1080px). The dots were 0.4 cm in diameter and the bars were 1.1 cm wide with varying height, but low bars were around 1 cm and high bars were around 4.5 cm tall. The visualizations were centered on a gray background [rgb: 128, 128, 128]. 

\textbf{Procedure.}
The instructions and procedure were the same as in Exps. 1 and 2. For the anchoring in the instructions, participants saw 10 example dot plots or bar charts (given their condition) that had a high-high and a low-low association palette for each emotion. 

\begin{figure*}[tb]
 \centering
 \includegraphics[width=.9\textwidth]{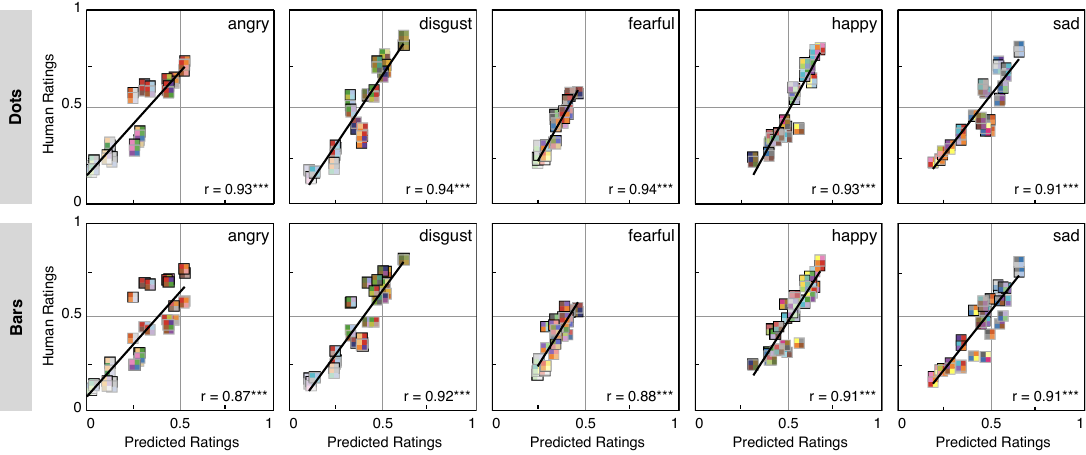}
 \caption{Scatterplots showing the mean human color-emotion associations with predicted association ratings (unweighted mean) within each chart condition (dot or bar). Palettes outlined in black represent ratings when the strong associate colors had a larger y-value, and palettes outlined in gray represent ratings when weak associate colors had a larger y-value. *$p < .05$, **$p < .01$, ***$p < .001$.}
 \label{fig:Exp3 Results}
 \vspace{-4mm}
\end{figure*}

\subsection{Results and discussion}

To test the additivity hypothesis, we investigated how well the mean emotion associations for the dot plots and bar charts were predicted by the simple, unweighted mean estimated emotion association of the colors within their palettes. For each of the 18 palettes for each emotion, we separated responses by which colors had greater y-values, but averaged over mirror versions of each plot (e.g., combining ascending/left-high with descending/right-high), resulting in 36 data points per emotion. As shown in Fig. \ref{fig:Exp3 Results} and Table \ref{tab:exp3_corr_comparisons}, the unweighted mean strongly predicted emotional associations of the whole visualizations for dot plots (\textit{r} ranging from .91 to .94) and bar charts (\textit{r} ranging from .81 to .92), supporting the additivity hypothesis. Histograms of correlations at the participant level can be found in Supplementary Fig. \ref{fig:Exp 3 Histograms}.

The next sequence of analyses tested the data-dependence hypothesis. We computed the weighted mean for each palette by multiplying the estimated association for each color by the y-value of the data mark assigned to that color. For bar charts, the y-value corresponded to the length of the bars, so it was a metric of region size. Thus, the data-dependence hypothesis implied that the weighted predictions would fit participant ratings better than the unweighted predictions would. For dot plots, the y-value corresponded only to height, with no impact on size.  If data-dependence concerned only size and not position, then weighted predictions should fit the data \textit{worse} than unweighted predictions because multiplying by the y-values would add non-informative noise. However, if data-dependence concerned position as well as size, then weighted predictions could be a better fit than unweighted predictions for dot plots. 

\begin{table}[H]
\small	

    \centering
    \caption{Correlations (\textit{r}) between mean participant emotion associations and the unweighted or weighted predictions (\textit{df} = 34 and all \textit{p}s <.001), with comparisons across correlations using the Hotelling-Williams test 
    (2-tailed because we hypothesized differences could go in either direction for dot plots).}
    
    \begin{tabular}{llcccc}
    \hline
         \textbf{Plot Type} & \textbf{Emotion} & \textbf{Unweighted \textit{r}} & \textbf{Weighted \textit{r}} &\textit{\textbf{t} }& \textit{\textbf{p}}\\
             \hline
         Dots & Angry  & .93 & .85 &  -2.44 & .020\\
   
        & Disgust   & .94 & .86 & -2.82 & .008\\
    
        & Fearful  & .94 & .86 & -2.65 & .012 \\
   
        & Happy  & .93 & .84 & -2.91 & .007 \\
  
         & Sad  & .91 & .78 &  -3.18 &  .003 \\
             \hline
        Bars & Angry  & .87 & .92 & 1.52 & .139 \\
  
        & Disgust   & .92 & .92 & -0.01 & .995\\
      
        & Fearful  & .88 & .87 & -0.20 & .846 \\
  
        & Happy  & .91 & .95 & 1.52 & .138\\
  
         & Sad  & .92 & .94 & 0.93 & .358\\
             \hline
    \end{tabular}

    \label{tab:exp3_corr_comparisons}
     \vspace{-2mm}
\end{table}

Table \ref{tab:exp3_corr_comparisons} shows the correlations between mean human ratings and predictions from the weighted vs. unweighted mean associations. We did not include scatterplots for the weighted predictions in Fig. \ref{fig:Exp3 Results} because they appear quite similar to the unweighted version. For bar charts, correlations for weighted predictions were numerically higher than for unweighted predictions in three out of five emotions but not significantly so (correlations were already so high there was not much room for improvement). For dot plots, correlations for weighted predictions were significantly lower than for unweighted predictions for all five emotions (Table \ref{tab:exp3_corr_comparisons}).   The weaker correlations using weighted predictions for dot plots provides evidence against position-based data-dependence for dots---weighting mean associations by y-value for dots \textit{decreased} model fits, suggesting that weighting by position added non-informative noise to the predictions.

\begin{figure}[tb]
 \centering
 \includegraphics[width=1.0\columnwidth]{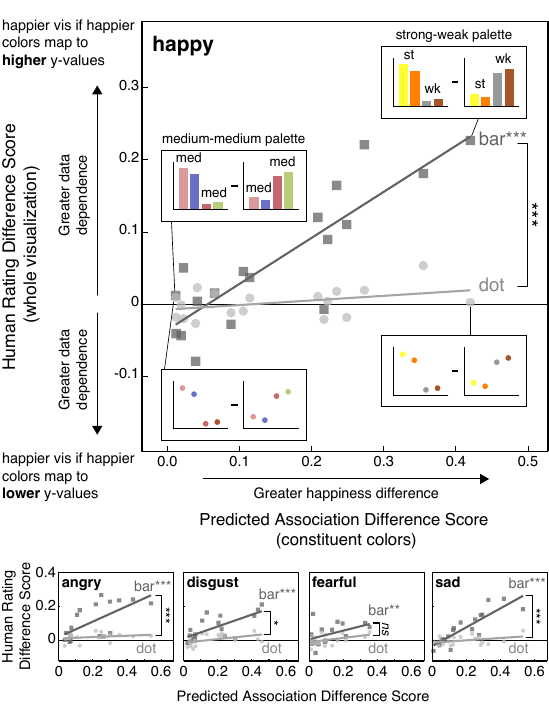}
 \caption{Scatterplots for each emotion showing the mean human rating difference scores for whole visualizations (y-axis) as a function of predicted association difference scores of constituent colors within each palette (x-axis), separated by dot plots (light gray circles) and bar charts (dark gray squares); see text for details. $*p<.05$, $**p<.01$, ***$p<.001$}
 
 \label{fig:exp3_diff}
 \vspace{-6mm}
\end{figure}

At first blush, it may seem that the effect of data dependence is weak (if existent) for bar charts, compared to the strong effects observed for shifted colormaps in Exp. 2. However, there is an important difference in how the affective connotations of colors varied within visualizations in Exp. 2 and Exp. 3. In Exp. 2, the endpoints of the color scales typically had different affective connotations (e.g., lighter endpoints were more positive), so shifting the underlying datasets changed the prevalence of colors of particular affective connotations. In Exp. 3, by design, some palettes contained colors with widely different affective connotations (e.g., strong--weak palettes) whereas others had similar affective connotation (e.g., medium-medium palettes) (Fig. \ref{fig:palettes}B, Fig. \ref{fig:exp3_diff}). The data-dependence hypothesis implies varying bar height should have a strong effect on affective connotation of the whole visualization when bar colors have different affective connotations, and should have little effect when bar colors have similar affective connotations.

To test this possibility, we computed `predicted association difference scores' as the difference of the mean estimated associations of the two most strongly associated colors minus the two most weakly associated colors within each palette (i.e., left two colors and right two colors within each palette, respectively, in Fig. \ref{fig:palettes}B). We then computed `human rating difference scores' for each visualization as the difference between mean association ratings for plots in which the stronger associated color pair had higher y-values minus ratings for plots in which the stronger associated color pair had the lower y-values. Fig. \ref{fig:exp3_diff} shows the relation between the predicted association difference scores and the human rating difference scores for each emotion (with separate points and best fit lines for bar charts and dot plots). We will walk through an example of how to interpret this figure prior to discussing the analyses. In the plot for happy (Fig. \ref{fig:exp3_diff}, top), the yellow/orange/gray/brown palette (strong--weak) has a large predicted association difference because yellow and orange are far happier than gray and brown, whereas the pink/blue/brown/green palette (medium--medium) has a low predicted difference score because the colors in the palettes have similar associations with happy. Aligning with the data-dependence hypothesis, the bar charts with the strong--weak palette had a positive rating difference score (i.e., was judged as happier when the yellow and orange bars had greater y-values than the gray and brown bars), whereas the bar charts with the medium--medium palette had a rating difference score near zero (i.e., was judged similarly regardless of bar height). This effect of association difference for bar charts did not appear to occur for dot plots, as varying the y-position did not affect dot size. 

These observations were supported by two sets of mixed-effects linear regression models. The first set tested whether predicted association difference had a larger effect (steeper slope in Fig. \ref{fig:exp3_diff}) for bar charts than for dot plots, which would occur if mark y-position affected associations for bar charts more than for dot plots. We included fixed effects for predicted association difference (centered), plot type (dot, bar), their interaction, and by-subject random intercepts and slopes for predicted association difference. Significant interactions indicated that predicted association difference indeed had a larger effect for bar charts than for dot plots for happy, angry, disgust, and sad, but not fearful (Table \ref{tab:exp3_diff_model}; see Supplementary Table \ref{tab:exp3_model_results} for full model output). The second set of models tested for effects predicted by association difference within each plot type (bar or dot only). Increasing association difference had a significant effect for bar charts and no significant effect for dot plots for all five emotions (Table \ref{tab:exp3_diff_model}; see Supplementary Table \ref{tab:exp3_plot_type_model_results} for full model output). These results indicate the size of the colored regions in bar charts (determined by y-values) had a greater influence on affective connotation as the difference in emotional associations of the larger vs. smaller colored regions increased. There was not evidence for such effects for dot plots constructed from the same underlying datasets because the y-values influenced only dot height, not size.

\begin{table}[t!]
\centering
\small	
\caption{Summary of mixed-effect models predicting human rating association differences from predicted association differences (Fig. \ref{fig:exp3_diff}). See Supplementary Tables \ref{tab:exp3_model_results}  and \ref{tab:exp3_plot_type_model_results} for full model output.}
\begin{tabular}{l ll ll ll}
\toprule
 & \multicolumn{2}{c}{Bar Only} & \multicolumn{2}{c}{Dot Only} & \multicolumn{2}{c}{Interaction} \\
\cmidrule(lr){2-3} \cmidrule(lr){4-5} \cmidrule(lr){6-7}
Emotion & $\beta$ & $p$ & $\beta$ & $p$  & $\beta$ & $p$  \\
\midrule
Angry   & 0.410 & $<.001$ & 0.046 & .228 & 0.364  & $<.001$  \\
Disgust & 0.327 & $<.001$ & 0.106 & .097 & 0.220  & .020  \\
Fearful & 0.170 & $.008$  & 0.094 & .058  & 0.076  & .331 \\
Happy   & 0.524 & $<.001$ & 0.068 & .090 & 0.456  & $<.001$  \\
Sad     & 0.515 & $<.001$ & 0.056 & .134 & 0.459  & $<.001$  \\
\bottomrule
\label{tab:exp3_diff_model}
\vspace{-3em}
\end{tabular}
\end{table}

\textbf{Summary.} Exp. 3 showed strong support for the additivity hypothesis---emotional associations for dot plots and bar charts were well predicted from the mean estimated associations of the individual colors. Our initial test of the data-dependence hypothesis comparing correlations using weighted vs. unweighted mean associations as predictors showed little evidence for data dependence for bar charts but suggested that data-dependence does not operate on mark position for dot plots, as weighting by y-values led to worse fits. However, when we accounted for the magnitude of variation in emotional association among colors within each palette, we found strong evidence for data dependence for bar charts. That is, emotional associations for bar charts were more strongly impacted by changes in y-values as colors assigned to high vs. low y-values had increasingly larger differences in emotional associations. Together, these results imply that designers need not be concerned with data dependence if all colors appear in equal amounts within a visualization (e.g., dot plots), and the extent to be concerned with effects of region size depends on how much the constituent colors differ in affective connotation.

\section{General Discussion and Conclusion}

Previous work showed that colors strongly influence affective connotation of visualizations and described ways in which dimensions of color modulate affective connotation \cite{bartram2017, braun2026, anderson2022}, but it was unclear how to use previous models to predict affective connotation of visualizations that varied widely in hue. Here, we addressed this gap by developing an approach that predicts emotional associations of whole visualizations using model estimated associations of the constituent colors. We tested the viability of our approach using continuous color scales in visualizations of continuous data (Exp. 1 and 2) and discrete color palettes in visualizations of categorical data (Exp. 3). 
We found that affective connotation of whole visualizations could be predicted by the aggregate of the emotional associations of the constituent colors (additivity hypothesis), but it was important to account for the size of colored regions when colored regions were disproportionately sized (data-dependence hypothesis), especially when the constituent colors varied in association strength with the given emotion. 

\begin{figure}[tb]
 \centering
 \includegraphics[width=0.9\columnwidth]{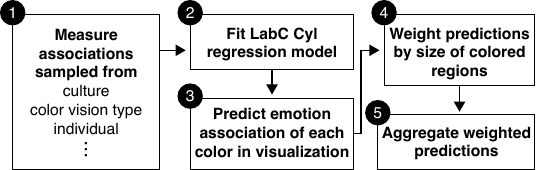}
 \caption{Steps to account for group or individual differences when predicting affective connotation of whole visualizations from constituent colors.  }
 \label{fig:indgroup}
 \vspace{-6mm}
\end{figure}

Our approach has potential to predict color-based affective connotation for a broad range of visualizations without having to collect new association data for individual colors, given the success of predictions generated from estimated color-emotion associations using pre-fit regression models (though we highlight limitations and open challenges below). The procedure is as follows (Fig. \ref{fig:process} and \ref{fig:indgroup}): (1) Extract RGB values from the visualization and translate to CIELCh color space. (2) For each unique color, use the regression coefficients in Table \ref{tab:reg_weights} and CIELCh value of the color to predict its emotional associations. (3) Weight the association of each color by the amount the color appears in the image (e.g., number of pixels, but we will consider potential problems with a pixel-based approach in the next section). (4) Aggregate the weighted values across colors in the image. Our results suggest this procedure is effective for predicting relative associations among visualizations, but additional scaling may be necessary to capture exact values, as discussed in Exp. 1. A designer could use our approach toward a variety of goals, including selecting colors that match a desired affective connotation, assessing the affective connotation of existing visualizations and flagging those with unintended emotional connotations, or comparing design alternatives. 

\textbf{Limitations and Challenges.} Although our approach was effective for the visualizations tested in the present study, several limitations and challenges need to be addressed before using our approach at scale. 

\textbf{\textit{Robustness of pixel-based weighting.}} A pixel-based approach to quantify the prevalence of each color in a visualization may be problematic if there are visual artifacts or ``hidden'' pixels, which are not visible to the human eye but bias predicted emotional associations. For example, consider two high resolution colormaps with the same exact pixels, structured so one colormap appears mostly shades dark red (strongly angry) with a light cyan (not angry) stripe down the middle, and the other distributes the light cyan pixels across the image so they are invisible to the naked eye. Our current approach would predict these two colormaps have the same estimated associations, but participants would likely judge the version with the perceivable light cyan stripe as less angry. A more robust approach to handling how spatial structure of pixels influences how colors are perceived is needed to automate predicting affective connotation of visualizations at scale. 

\textbf{\textit{Data-dependence based on mark frequency.}} Our comparison of bar charts and dot plots suggests that region size, not position, drives data-dependence, but questions remain about how data-dependence plays out in other types of visualizations. For example, in multi-class scatterplots, would data dependence play out in the total visible area of each color combined over distinct marks? If there is partial occlusion due to overplotting, should the estimates account for the visible colors at the pixel level (image-based representation), or the perceived amount of each color after accounting for amodal completion behind partially occluded colored regions (surface-based representation)? Research on effects of relative size on preference for color combinations suggests that the surface-based representation may dominate \cite{schloss2011role}, but given other differences between color preferences and emotional associations highlighted in Section \ref{sec:additivity}, this hypothesis needs to be tested.

\textbf{\textit{Visualization component relevance.}} Automating our approach will require extracting colors from ``relevant'' components of visualizations and ignoring ``irrelevant'' components, but doing so requires knowing which components are relevant.  For example, do the colors of axes, grid lines, mark outlines, text, or the background contribute to affective connotation, or only marks that represent data (as assumed in the present study)? Careful testing will be needed to answer this question. 

\textbf{\textit{Group and individual variation.}} In this study, our estimates of color-emotion associations of individual colors were computed based on data aggregated from undergraduate participants with typical color vision in the United States. These estimates were effective for predicting emotional associations of whole visualizations judged by participants from the same population.  However, there should be group (e.g., culture, color vision type) and individual differences in emotional associations for whole visualizations to the extent to which there are group/individual differences in emotional associations for constituent colors in the visualizations (see \cite{jonauskaite2025, ou2012, gao2007, tham2020systematic} for discussions of cultural similarities/differences). Our approach suggests that such group/individual differences should be predictable by measuring color-emotion associations sampled from the group/individual of interest (step 1 in Fig. \ref{fig:indgroup}) and then repeating our procedure in this paper (steps 2-5 in Fig. \ref{fig:indgroup}), using the association data from the group or individual of interest. This logic assumes that additivity and data dependence work similarly across groups and individuals as they did in the present study, and this assumption will need to be tested empirically.  

\textbf{\textit{Other design factors.}} Many other factors beyond color contribute to affective connotation of visualizations \cite{lan2024}. Further experimentation and modeling will be needed to predict how color combines with other factors to determine the affective connotations of whole visualizations. 

\textbf{Conclusion.} We developed and tested an approach to predict affective connotation of whole visualizations from the affective connotation of the constituent colors. This approach uses pre-fit regression models to estimate color-emotion associations for individual colors, and aggregates those values to compute predicted associations for whole visualizations. Our results showed that this approach is effective for predicting human judgments of emotional associations for colormap data visualizations, dot plots, and bar charts, with potential to scale to automatically predict affective connotation for a wider variety of visualization types. This work can be used to help design visualizations that convey specific affective connotation to support visual communication. 

\section*{Use of AI} 
GPT-5.3 and Claude 4.8 Opus were used to help write web experiment code, format tables, and debug image compression issues.

\bibliographystyle{abbrv-doi-hyperref-narrow}

\bibliography{template}

\clearpage
\appendix 
\renewcommand*{\thesection}{S}
\counterwithin{figure}{section}
\counterwithin{table}{section}

\section{Supplementary Material}\label{sec:supplementary}

\subsection{Selecting underlying dataset} \label{sec:underlyingdata}
\begin{figure*}[t!]
 \centering
 \includegraphics[width=\textwidth]{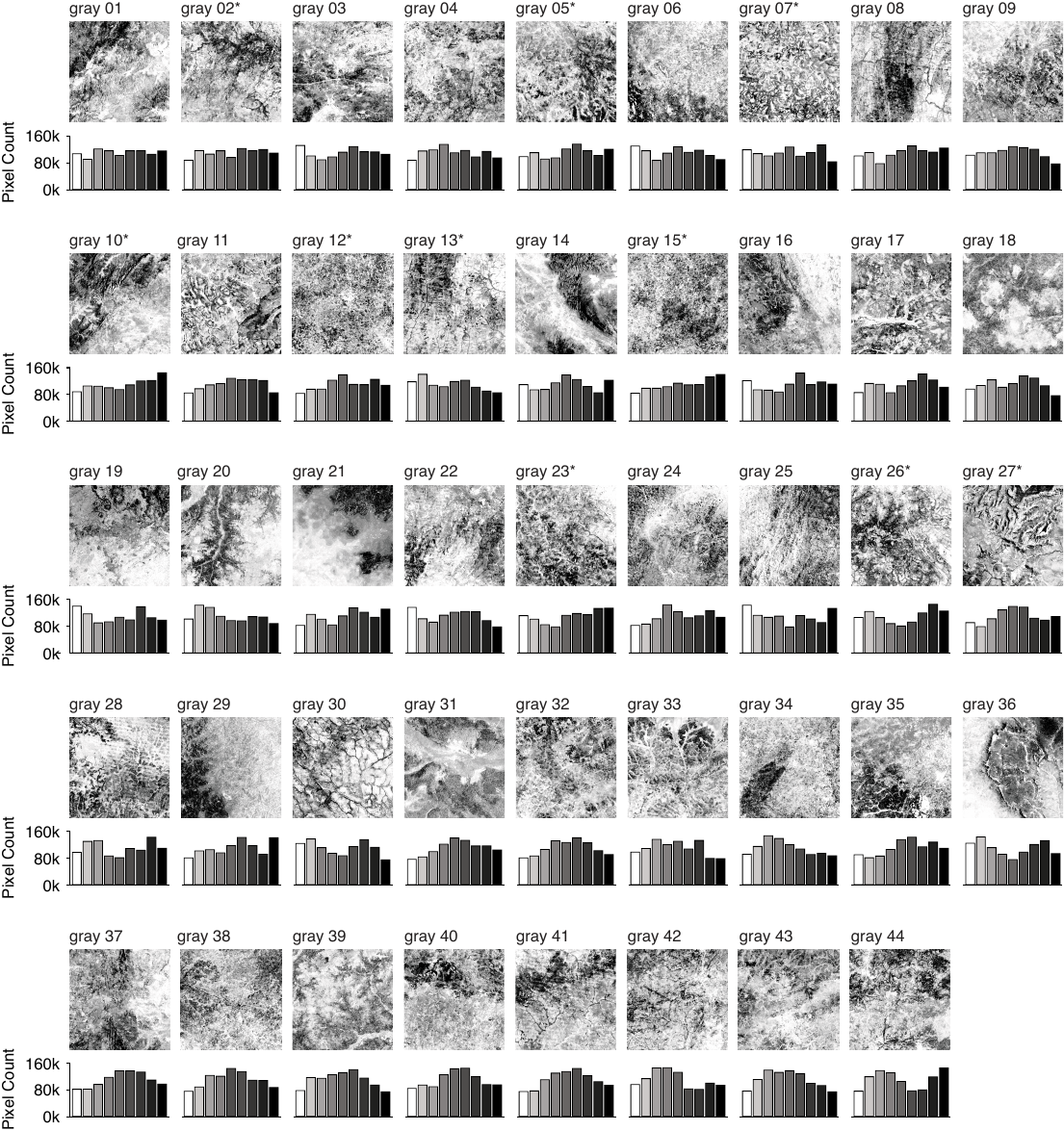}
 \caption{Grayscale colormaps of the 44 subtiles (figure reproduced from \cite{braun2026}). The colormaps show biomass density values scaled to range from 0 to 100, grouped into 9 bins, and then mapped to a 9-step gray color scale. The histogram below each colormap shows the number of pixels in each of the 9 bins, with each bar color in the histogram corresponding to the pixel colors assigned to values in that bar's bin. The maps are labeled 01 to 44 in order from most to least uniform distribution within the set. An asterisk next to a map number indicates that map was selected for Exps. 1 and 2.}
 \label{fig:graymaps44}
 \vspace{-3mm}
\end{figure*}

This section describes Braun et al.’s approach to select the underlying datasets (text reproduced from  \cite{braun2026}):
 
 \begin{quote}
 Each pixel in the dataset has a spatial resolution of 30$\times$30 meters and represents megagrams of AGB per hectare (Mg ha$^{-1}$) in the year 2000. The dataset includes 280 `tiles' across the world (40,000$\times$40,000 pixels per tile). We split each tile into smaller 1,000$\times$1,000 pixel `subtiles' (448,000 subtiles total). We then excluded subtiles that had N/A values, corresponding to 0 Mg ha$^{-1}$, to avoid including subtiles that had easily recognizable geographical features (e.g., rivers or croplands). 

Of the remaining set, we sought to find subtiles that had the most uniform distribution of values, such that when we applied a color scale to produce a colormap, all colors in the color scale would be well-represented in the map. Although we could have generated synthetic datasets that were perfectly uniform, we used real-world data for ecological validity. We used the following steps to filter the subtiles. First, we rescaled the values from units of biomass density to a standardized scale from 0 to 100. Second, we eliminated subtiles whose medians were outside of a range of 40-60 to avoid skewed distributions. Third, for each of the subtiles we grouped the values into nine bins corresponding to the nine colors within the Color Crafter color scales and computed the pixel count for each bin. We kept subtiles for which the pixel counts of each of the nine bins fell within an acceptable range of deviation from the expected pixel count for each bin under a uniform distribution (1000$\times$1000 pixels/9 bins).\footnote{We operationalized `acceptable' as within ±1/3 of the expected bin counts. We explored other metrics to assess deviation from a uniform distribution, but we found this approach was most effective for ensuring that no bin had an outsized number of pixels.} This method resulted in 44 maps (Fig. \ref{fig:graymaps44}).

\end{quote}

\subsection{Creating shifted datasets}
\label{sec:supp-shift}
This section describes Braun et al.’s \cite{braun2026} approach to shifting datasets:

    To create the shifted versions, they first normalized the values in each subtile (0--1; min-max normalization). Next, they created `light shifted' maps (i.e., a greater proportion of lighter values) by applying the following formula to each pixel $x$ in the data matrix

$$
b - \frac{(b - x)^p}{(b - a)^{p - 1}}
$$

where $a$ and $b$ were the limits of allowable data values (0 and 1, respectively). The parameter $p$ determined the degree of shift (larger values of $p$ resulting in larger shifts). 

The analogous equation for making `dark shifted' colormaps was  

$$
\frac{(x - a)^{p}}{(b - a)^{p - 1}} + a
,$$

\begin{figure}[ht!]
 \centering
 \includegraphics[width=1\columnwidth]{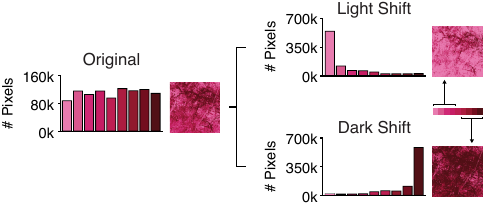}
 \caption{Diagram from \cite{braun2026} of how the underlying dataset for a colormap was light shifted and dark shifted. The histograms for each colormap show the number of pixels in each color from the color scale used to produce the colormap (bar colors correspond to pixel colors in colormaps).}
 \label{fig:ShiftingColormaps}
 \vspace{-4mm}
\end{figure}

They set the value of $p$ to be 4 after visually inspecting the light and dark shifted maps for a range of parameter values. This parameter value resulted in shifted maps that were perceptually distinct from the non-shifted version while preserving enough visual detail that they still appeared to be generated from the same data distribution (Fig. \ref{fig:ShiftingColormaps}).

\subsection{Color space regression}

As described in the main text (Section \ref{sec:color_space_regression}), the LabC Cyl2 model can be used to characterize patterns of color-emotion associations across colors. Table \ref{tab:reg_weights} shows the regression weights for each parameter (Intercept, L, C, sin(H), cos(H), sin(2H), cos(2H)) for each of the five emotions tested in the recent study (angry, disgust, fearful, happy, and sad).

\begin{table}[H]
\small	
    \centering
    \caption{Weights ($w_1 - w_7$) on each of the parameters in the color space regression equation for each emotion (computed using color-emotion association data for the UW-71 colors reported in Mukherjee et al. \cite{mukherjee_llm})}
    \begin{tabular}{lcccccc}
    \hline
        \textbf{Parameter} &	\textbf{Weight}  & \textbf{Angry}     & \textbf{Disgust}   & \textbf{Fearful}   & \textbf{Happy}     & \textbf{Sad} \\
        \hline 
        Intercept &	$w_1$	    & 0.5258    & 0.6516    & 0.6305    & -0.0154	& 0.8559\\
        L	      & $w_2$	    &-0.0048	& -0.0044	& -0.0039	& 0.0057	& -0.0044 \\
        C	      & $w_3$	    & 0.0013	& -0.0009	&-0.0012	& 0.0041	& -0.0037 \\
        sin(H)	  & $w_4$	    & 0.0992	& 0.2007	& 0.0458	& -0.0913	& -0.0692 \\
        cos(H)	  & $w_5$       & 0.0628	& -0.0402	& 0.0222	& -0.0063	& -0.0578 \\
        sin(2H)	  & $w_6$	    & 0.0917	& -0.0536	& 0.0112	& 0.0367	& -0.0029 \\
        cos(2H)	  & $w_7$	    & 0.0299	& -0.0428	& -0.0132	& 0.0392	& -0.0599\\
        \hline
\end{tabular}

    \label{tab:reg_weights}
     \vspace{-2mm}
\end{table}

\subsection{Sub-sampling color scales for Experiment 2} \label{sec:sub-sampling}

In Exp. 2, we reduced the total number of color scales from 40 to 20 in order to preserve the total number of trials while introducing the two new `shift' conditions for the underlying map datasets.
In doing so, we sought to find the set of 20 scales that were maximally dissimilar in terms of their mean association ratings with the five emotion terms. First, we represented each scale as a 5-dimensional vector, where each dimension was the mean association rating with one of the five emotions, and computed the scale $\times$ scale distance matrix by computing the pairwise Euclidean distance between each scale.
Next, we iteratively applied a greedy farthest-point sampling algorithm laid out in Algorithm \ref{alg:fps}. The goal was to arrive at a set of 20 color scales, $\mathcal{D}$, sampled from the full set of scales 40 scales, $\mathcal{S}$. We begin by finding the scales $i^*$ and $j^*$ that are maximally distant in the distance matrix, $D$. Next, for each of the remaining 38 scales we found their nearest neighbor among the selected scales and added the scale that had the \textit{maximum} nearest-neighbor distance. We iteratively continued this process until our selected set had 20 color scales.
\begin{algorithm}
\caption{Subsampling maximally dissimilar color scales}\label{alg:fps}
\begin{algorithmic}[1]

\Ensure Subset $\mathcal{D} \subseteq \mathcal{S}$ with $|\mathcal{D}| = 20$
\State Compute Euclidean distance matrix $\mathbf{D}$ where $D_{ij} = \|s_i - s_j\|_2$
\State $(i^*, j^*) \leftarrow \arg\max_{i,j} D_{ij}$ \Comment{maximally different pair of scales}
\State $\mathcal{D} \leftarrow \{s_{i^*}, s_{j^*}\}$
\While{$|\mathcal{D}| < 20$}
    \State $\mathcal{R} \leftarrow \mathcal{S} \setminus \mathcal{D}$
    \For{each $s_r \in \mathcal{R}$}
        \State $\delta(s_r) \leftarrow \min_{s_d \in \mathcal{D}} D_{r,d}$
    \EndFor
    \State $s^* \leftarrow \arg\max_{s_r \in \mathcal{R}} \delta(s_r)$ \Comment{distance between non-selected scale and nearest selected scale}
    \State $\mathcal{D} \leftarrow \mathcal{D} \cup \{s^*\}$  \Comment{Adding furthest non-selected scale}
\EndWhile
\State \Return $\mathcal{D}$
\end{algorithmic}

\end{algorithm}

\subsection{UW-71 Colors} \label{sec:UW71}
\begin{figure}[h!]
 \centering
 \includegraphics[width=\columnwidth]{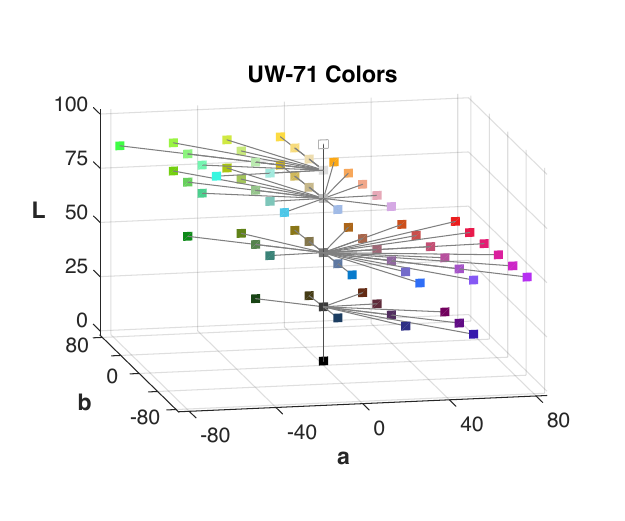}
 \caption{The UW-71 colors plotted in CIELAB color space (from \cite{mukherjee2022}).}
 \label{fig:UW_71}
 \vspace{-4mm}
\end{figure}

The regression models used to generate the predictions for this study were fit using color-emotion association data for the ``UW-71 colors'' (Fig. \ref{fig:UW_71}). As described in Mukherjee et al. \cite{mukherjee2022}, the UW-71 colors are based on the UW-58 colors described in \cite{rathore2020}, but extended to include lighter yellows and greens of higher saturation. The UW-58 colors includes 58 colors uniformly sampled over CIELAB space, with an edge distance of $\Delta$E = 25. The grid was rotated around the L* axis by 3 degrees to increase the number of colors, as described in \cite{rathore2020}). The UW-71 include an additional 13 colors by sampling an additional plane lightness L* = 88, maintaining $\Delta$E = 25 within the plane. 

\subsection{Unweighted mean vs. weighted mean models in Exp. 2}
Table S.2 shows the full results of mixed-effects linear regression models predicting per-trial ratings from unweighted mean association and weighted mean association predictors. Table \ref{tab:exp2-uniform-v-weighted} in the main text shows that the weighted models consistently fit the human ratings better than the unweighted models.

\begin{table}[h!]
\centering
\small	
\caption{Mixed-effects linear regression results for each emotion in Exp. 2, predicting mean human ratings from the unweighted vs. weighted mean association ratings (Assoc.) of the constituent colors.}

\begin{tabular}{lllccccc}
\toprule
Model &Emotion & Predictor & $\beta$ & SE & df & $t$ & $p$ \\
\midrule
Unweighted &Angry & Intercept & 0.448 & 0.019 & 39.0 & 23.83 & $< .001$ \\
& & Assoc.  & 0.859 & 0.079 & 39.0 & 10.91 & $< .001$ \\
\addlinespace
&Fearful & Intercept & 0.459 & 0.017 & 39.0 & 27.77 & $< .001$ \\
& &  Assoc. & 0.694 & 0.147 & 39.0 & 4.71 & $< .001$ \\
\addlinespace
&Happy & Intercept & 0.441 & 0.014 & 39.0 & 32.40 & $< .001$ \\
& &  Assoc. & 0.436 & 0.084 & 39.0 & 5.17 & $< .001$ \\
\addlinespace
&Sad & Intercept & 0.415 & 0.021 & 39.0 & 20.09 & $< .001$ \\
& &  Assoc. & 0.807 & 0.084 & 39.0 & 9.57 & $< .001$ \\
\addlinespace
&Disgust & Intercept & 0.427 & 0.017 & 39.0 & 24.91 & $< .001$ \\
& &  Assoc. & 0.873 & 0.086 & 39.0 & 10.21 & $< .001$ \\

\midrule

Weighted & Angry & Intercept & 0.448 & 0.019 & 39.0 & 23.85 & $< .001$ \\
& &  Pred. & 0.896 & 0.069 & 39.0 & 13.07 & $< .001$ \\
\addlinespace
&Fearful & Intercept & 0.459 & 0.017 & 39.0 & 27.81 & $< .001$ \\
& &  Pred. & 1.094 & 0.140 & 39.1 & 7.80 & $< .001$ \\
\addlinespace
&Happy & Intercept & 0.441 & 0.014 & 39.0 & 32.40 & $< .001$ \\
& &  Pred. & 0.849 & 0.092 & 39.0 & 9.20 & $< .001$ \\
\addlinespace
&Sad & Intercept & 0.415 & 0.021 & 39.0 & 20.11 & $< .001$ \\
& &  Pred. & 0.962 & 0.093 & 39.0 & 10.31 & $< .001$ \\
\addlinespace
&Disgust & Intercept & 0.427 & 0.017 & 39.0 & 24.92 & $< .001$ \\
& &  Pred. & 0.685 & 0.072 & 38.9 & 9.56 & $< .001$ \\
\bottomrule
\label{tab:supp-exp2-models}
\end{tabular}
\vspace{-3em}
\end{table}

\FloatBarrier

\subsection{Full model results for Exp. 3} \label{sec:exp3-model}
Supplementary Table \ref{tab:exp3_model_results} shows the full output of the linear mixed-effects regression models predicting human rating difference scores from the predicted association difference scores, plot type (bar vs. dot) and their interaction). Positive $\beta$ values for interactions indicate a greater effect for bar plots than dot plots. Supplementary Table \ref{tab:exp3_plot_type_model_results} shows the full output of the corresponding models conducted within each plot type. These results correspond to the summary in Table \ref{tab:exp3_diff_model} of the main text.

\begin{table}[h!]
\small	

   \centering
   \caption{Mixed-effects model results from Exp. 3, predicting human rating difference scores from predicted association difference scores, plot type, and their interaction for each emotion.}
   \begin{tabular}{llllcccc}
   \hline
        \textbf{Emotion} & \textbf{Predictor} & \textbf{$\beta$} & \textbf{df} & \textbf{SE} & \textbf{$t$} & \textbf{$p$}  \\
            \hline
        Angry  & Intercept & 0.070 & 77.00 & 0.009 & 7.749 & <.001 \\
               & Assoc. Diff. & 0.228 & 77.01 & 0.032 & 7.156 & <.001 \\
               & Plot Type & 0.103 & 77.00 & 0.018 & 5.672 & <.001 \\
               & Assoc. Diff.:Plot & 0.364 & 77.01 & 0.064 & 5.707 & <.001 \\
           \hline
       Disgust & Intercept & 0.033 & 77.00 & 0.006 & 5.175 & <.001 \\
               & Assoc. Diff. & 0.217 & 77.00 & 0.046 & 4.660 & <.001 \\
               & Plot  & 0.063 & 77.00 & 0.013 & 4.873 & <.001 \\
               & Assoc. Diff.:Plot & 0.220 & 77.00 & 0.093 & 2.371 & .020 \\
           \hline
       Fearful & Intercept & 0.023 & 77.00 & 0.007 & 3.196 & .002 \\
               & Assoc. Diff. & 0.132 & 77.00 & 0.039 & 3.386 & .001 \\
               & Plot Type & 0.034 & 77.00 & 0.014 & 2.407 & .018 \\
               & Assoc. Diff.:Plot & 0.076 & 77.00 & 0.078 & 0.979 & .331 \\
           \hline
       Happy   & Intercept & 0.030 & 77.00 & 0.006 & 5.336 & <.001 \\
               & Assoc. Diff. & 0.296 & 77.01 & 0.038 & 7.696 & <.001 \\
               & Plot Type & 0.056 & 77.00 & 0.011 & 4.929 & <.001 \\
               & Assoc. Diff.:Plot & 0.456 & 77.01 & 0.077 & 5.925 & <.001 \\
           \hline
       Sad     & Intercept & 0.035 & 77.01 & 0.007 & 5.231 & <.001 \\
               & Assoc. Diff. & 0.285 & 77.01 & 0.043 & 6.591 & <.001 \\
               & Plot Type & 0.078 & 77.01 & 0.013 & 5.882 & <.001 \\
               & Assoc. Diff.:Plot & 0.459 & 77.01 & 0.087 & 5.302 & <.001 \\
           \hline

   \end{tabular}

   \label{tab:exp3_model_results}
    \vspace{-3mm}
\end{table}

\begin{table}[h!]
\small	

   \centering
   \caption{Mixed-effects model results from Exp. 3, predicting human rating difference scores from predicted association difference scores for each emotion, separated by plot type.}
   \begin{tabular}{llllcccc}
   \hline
        \textbf{Plot Type} & \textbf{Emotion} & \textbf{Predictor} & \textbf{$\beta$} & \textbf{df} & \textbf{SE} & \textbf{$t$} & \textbf{$p$}  \\
            \hline
    Dots & Angry  & Intercept  & 0.019 & 38.00 & 0.008 & 2.358 & .024 \\
     &        & Assoc. Diff.  & 0.046 & 38.00 & 0.038 & 1.226 & .228 \\
\cline{2-8}
     & Disgust & Intercept  & 0.002 & 38.00 & 0.007 & 0.297 & .768 \\
     &         & Assoc. Diff.  & 0.106 & 38.00 & 0.062 & 1.704 & .097 \\
\cline{2-8}
     & Fearful & Intercept  & 0.006 & 38.00 & 0.007 & 0.785 & .437 \\
     &         & Assoc. Diff.  & 0.094 & 38.00 & 0.048 & 1.952 & .058 \\
\cline{2-8}
     & Happy   & Intercept  & 0.002 & 38.00 & 0.007 & 0.336 & .739 \\
     &         & Assoc. Diff.  & 0.068 & 38.00 & 0.039 & 1.739 & .090 \\
\cline{2-8}
     & Sad  & Intercept  & -0.004 & 38.00 & 0.005 & -0.803 & .427 \\
     &      & Assoc. Diff. & 0.056 & 38.00 & 0.036 & 1.532 & .134 \\
\hline
Bars & Angry & Intercept  & 0.122 & 39.00 & 0.016 & 7.548 & <.001 \\
     &       & Assoc. Diff. & 0.410 & 39.00 & 0.051 & 8.024 & <.001 \\
\cline{2-8}
     & Disgust & Intercept  & 0.065 & 39.00 & 0.011 & 5.886 & <.001 \\
     &         & Assoc. Diff. & 0.327 & 39.01 & 0.069 & 4.757 & <.001 \\
\cline{2-8}
     & Fearful & Intercept  & 0.040 & 39.00 & 0.012 & 3.268 & .002 \\
     &         & Assoc. Diff. & 0.170 & 39.00 & 0.061 & 2.786 & .008 \\
\cline{2-8}
     & Happy & Intercept  & 0.058 & 39.00 & 0.009 & 6.486 & <.001 \\
     &       & Assoc. Diff. & 0.524 & 38.99 & 0.066 & 7.980 & <.001 \\
\cline{2-8}
     & Sad & Intercept  & 0.074 & 39.00 & 0.012 & 6.140 & <.001 \\
     &     & Assoc. Diff. & 0.515 & 39.00 & 0.078 & 6.621 & <.001 \\
\hline
   \end{tabular}

   \label{tab:exp3_plot_type_model_results}

\end{table}


\begin{figure*}[h!]
 \centering
 \includegraphics[width=\textwidth]{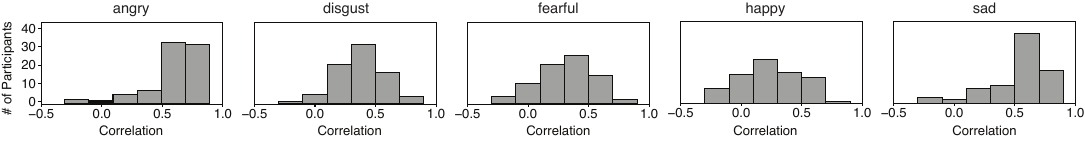}
 \caption{Histograms showing the frequency of Pearson's \textit{r} values between model predictions and individual participant ratings in Exp. 1. Correlations were computed separately for each emotion within each participant. Correlations were grouped into bins of width 0.2 for plotting. Given that each participant judged only one colormap data visualization for each of the 40 color scale $\times$ each of the two mappings (default and reversed) this histogram show correlations for data that neither aggregated over participants nor over colormap visualizations.  }
 \label{fig:Exp 1 Histograms}
 \vspace{-0mm}
\end{figure*}

\begin{figure*}[t!]
 \centering
 \includegraphics[width=\textwidth]{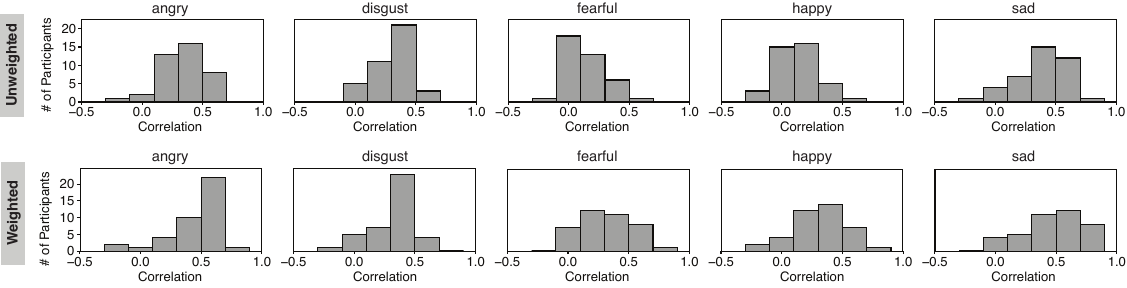}
 \caption{Histograms showing the frequency of Pearson's \textit{r} values between model predictions and individual participant ratings in Exp. 2. Correlations were computed separately for the unweighted (top row) and weighted by pixel count (bottom row) predicted ratings for each emotion within each participant. Correlations were grouped into bins of width 0.2 for plotting. As described in the main text, color scales in the default mapping applied to high-shifted datasets were analogous to color scales in the reversed mapping applied to the low-shifted datasets, so we averaged over these conditions. Likewise, we averaged over conditions that applied color scales in the reversed mapping to high-shifted datasets and color scales in the default mapping to the low-shifted datasets.
 }
 \label{fig:Exp 2 Histograms}
 \vspace{-0mm}
\end{figure*}

\begin{figure*}[t!]
 \centering
 \includegraphics[width=\textwidth]{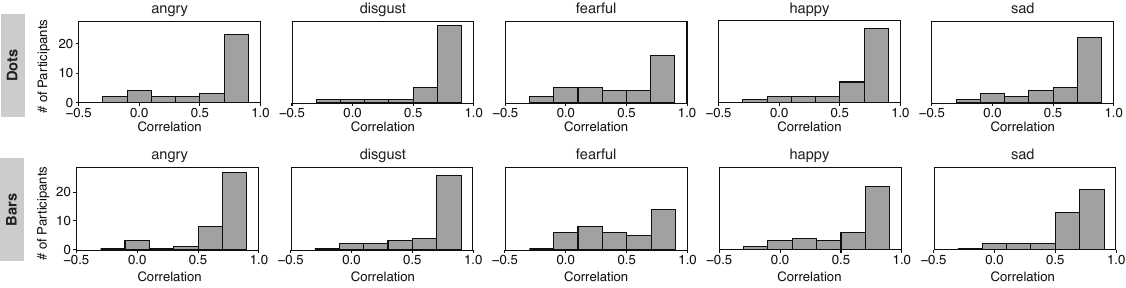}
 \caption{Histograms showing the frequency of Pearson's \textit{r} values between model predictions and individual participant ratings in Exp. 3. Correlations were computed separately for each emotion within each participant in the dot plot group (top row) and the bar chart group (bottom row). Each participant judged 72 visualizations per emotion in their assigned plot type (6 emotion strength pairings $\times$ 3 versions of each pairing $\times$ 2 association orders  $\times$ 2 y-value positions). Consistent with the analysis done for the correlations computed across participants, we again aggregated over mirrored versions of each visualization (e.g., combining ascending/left-high with descending/right-high) resulting in 36 data points per emotion per participant. Correlations were grouped into bins of width 0.2 for plotting.}
 \label{fig:Exp 3 Histograms}
 \vspace{-0mm}
\end{figure*}

\begin{figure*}
    \centering
    \includegraphics[width=1\linewidth]{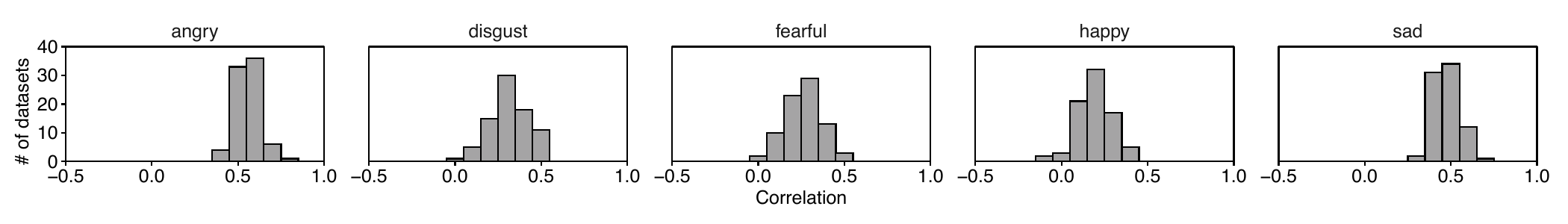}
    \caption{Histograms showing the frequency of Pearson's \textit{r} values between model predictions and participant ratings across all 80 color scales for each dataset in Experiment 1. That is, for each of the 80 underlying datasets (10 maps $\times$ 4 rotations $\times$ 2 mirror flips), we computed the correlation between model predictions and participant ratings when that dataset was depicted in each of the 80 color scales (40 named scales $\times$ 2 mappings). These correlation are at the individual visualization and participant level computed on raw ratings not aggregated over participants or visualizations.}
    \label{fig:dataset_cor_histograms}
\end{figure*}

\FloatBarrier
\begin{table}[t!]
\centering
\small
\caption{Mean association ratings between emotion concepts and colormaps constructed from each color scale in Exp.\ 1.}
\begin{tabular}{l c rrrrr}
\toprule
Color Scale & & angry & disgust & fearful & happy & sad \\
\midrule
Blues & \includegraphics[height=1.6ex,width=1.1cm]{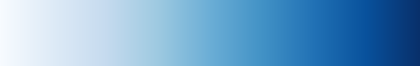} & 0.214 & 0.213 & 0.336 & 0.445 & 0.756 \\
BrBG & \includegraphics[height=1.6ex,width=1.1cm]{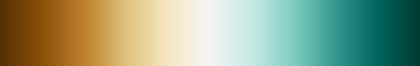} & 0.337 & 0.566 & 0.448 & 0.447 & 0.496 \\
BuGn & \includegraphics[height=1.6ex,width=1.1cm]{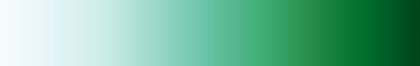} & 0.249 & 0.601 & 0.362 & 0.5 & 0.381 \\
BuPu & \includegraphics[height=1.6ex,width=1.1cm]{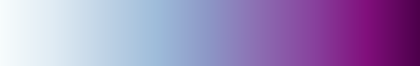} & 0.308 & 0.331 & 0.516 & 0.385 & 0.662 \\
GnBu & \includegraphics[height=1.6ex,width=1.1cm]{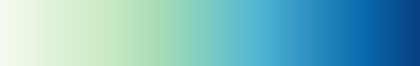} & 0.219 & 0.247 & 0.302 & 0.503 & 0.699 \\
Greens & \includegraphics[height=1.6ex,width=1.1cm]{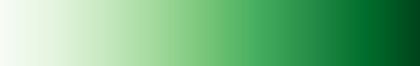} & 0.25 & 0.603 & 0.337 & 0.476 & 0.33 \\
Greys & \includegraphics[height=1.6ex,width=1.1cm]{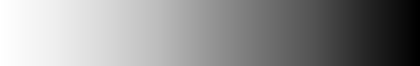} & 0.417 & 0.35 & 0.619 & 0.155 & 0.759 \\
OrRd & \includegraphics[height=1.6ex,width=1.1cm]{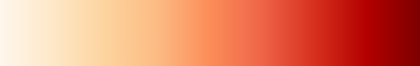} & 0.811 & 0.416 & 0.569 & 0.375 & 0.294 \\
Oranges & \includegraphics[height=1.6ex,width=1.1cm]{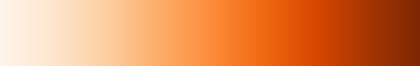} & 0.7 & 0.42 & 0.523 & 0.442 & 0.273 \\
PRGn & \includegraphics[height=1.6ex,width=1.1cm]{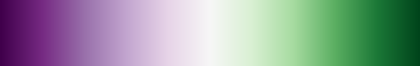} & 0.334 & 0.671 & 0.512 & 0.375 & 0.366 \\
PiYG & \includegraphics[height=1.6ex,width=1.1cm]{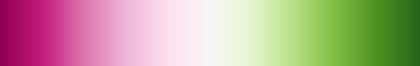} & 0.332 & 0.639 & 0.46 & 0.534 & 0.278 \\
PuBu & \includegraphics[height=1.6ex,width=1.1cm]{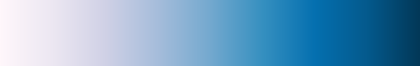} & 0.202 & 0.229 & 0.325 & 0.455 & 0.758 \\
PuBuGn & \includegraphics[height=1.6ex,width=1.1cm]{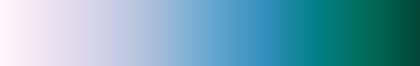} & 0.214 & 0.293 & 0.347 & 0.448 & 0.725 \\
PuOr & \includegraphics[height=1.6ex,width=1.1cm]{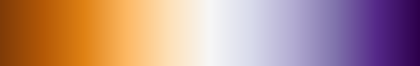} & 0.429 & 0.501 & 0.511 & 0.421 & 0.379 \\
PuRd & \includegraphics[height=1.6ex,width=1.1cm]{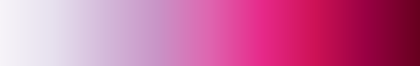} & 0.478 & 0.29 & 0.364 & 0.687 & 0.226 \\
Purples & \includegraphics[height=1.6ex,width=1.1cm]{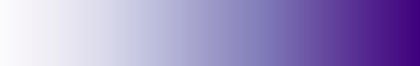} & 0.268 & 0.3 & 0.475 & 0.436 & 0.632 \\
RdBu & \includegraphics[height=1.6ex,width=1.1cm]{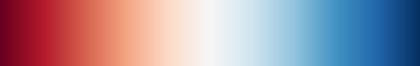} & 0.59 & 0.422 & 0.546 & 0.401 & 0.516 \\
RdGy & \includegraphics[height=1.6ex,width=1.1cm]{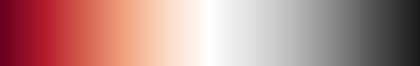} & 0.693 & 0.524 & 0.684 & 0.229 & 0.474 \\
RdPu & \includegraphics[height=1.6ex,width=1.1cm]{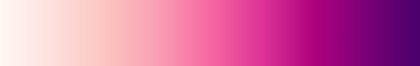} & 0.48 & 0.312 & 0.453 & 0.654 & 0.27 \\
RdYlBu & \includegraphics[height=1.6ex,width=1.1cm]{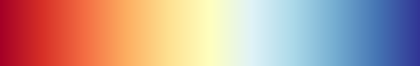} & 0.587 & 0.472 & 0.554 & 0.469 & 0.469 \\
RdYlGn & \includegraphics[height=1.6ex,width=1.1cm]{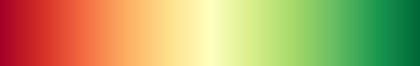} & 0.62 & 0.699 & 0.526 & 0.403 & 0.322 \\
Reds & \includegraphics[height=1.6ex,width=1.1cm]{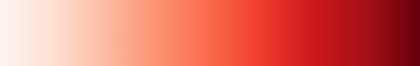} & 0.851 & 0.406 & 0.596 & 0.349 & 0.297 \\
Spectral & \includegraphics[height=1.6ex,width=1.1cm]{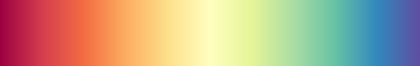} & 0.6 & 0.498 & 0.523 & 0.465 & 0.402 \\
YlGn & \includegraphics[height=1.6ex,width=1.1cm]{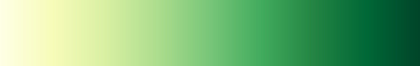} & 0.249 & 0.627 & 0.368 & 0.488 & 0.312 \\
YlGnBu & \includegraphics[height=1.6ex,width=1.1cm]{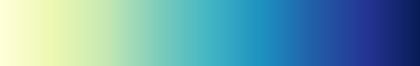} & 0.226 & 0.276 & 0.347 & 0.462 & 0.724 \\
YlOrBr & \includegraphics[height=1.6ex,width=1.1cm]{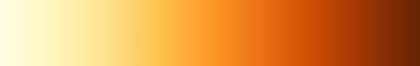} & 0.644 & 0.411 & 0.498 & 0.505 & 0.287 \\
YlOrRd & \includegraphics[height=1.6ex,width=1.1cm]{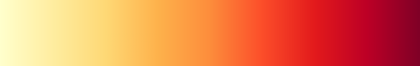} & 0.82 & 0.419 & 0.564 & 0.428 & 0.278 \\
berlin & \includegraphics[height=1.6ex,width=1.1cm]{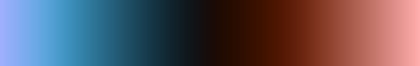} & 0.604 & 0.458 & 0.647 & 0.294 & 0.568 \\
bwr & \includegraphics[height=1.6ex,width=1.1cm]{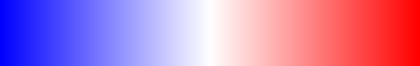} & 0.616 & 0.438 & 0.55 & 0.476 & 0.419 \\
cividis & \includegraphics[height=1.6ex,width=1.1cm]{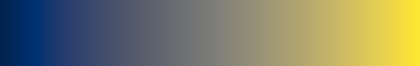} & 0.391 & 0.485 & 0.559 & 0.288 & 0.621 \\
coolwarm & \includegraphics[height=1.6ex,width=1.1cm]{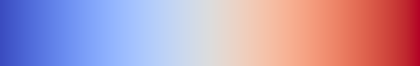} & 0.559 & 0.418 & 0.518 & 0.489 & 0.471 \\
gist\_earth & \includegraphics[height=1.6ex,width=1.1cm]{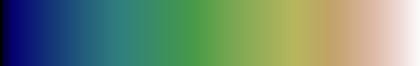} & 0.287 & 0.638 & 0.427 & 0.43 & 0.462 \\
inferno & \includegraphics[height=1.6ex,width=1.1cm]{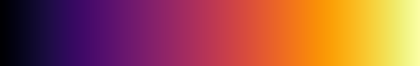} & 0.748 & 0.467 & 0.666 & 0.42 & 0.334 \\
magma & \includegraphics[height=1.6ex,width=1.1cm]{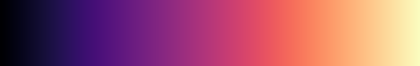} & 0.664 & 0.406 & 0.652 & 0.389 & 0.361 \\
managua & \includegraphics[height=1.6ex,width=1.1cm]{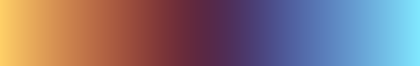} & 0.512 & 0.425 & 0.575 & 0.391 & 0.545 \\
plasma & \includegraphics[height=1.6ex,width=1.1cm]{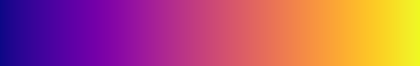} & 0.607 & 0.44 & 0.595 & 0.484 & 0.298 \\
seismic & \includegraphics[height=1.6ex,width=1.1cm]{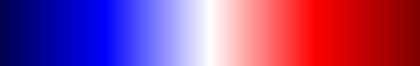} & 0.685 & 0.454 & 0.609 & 0.423 & 0.431 \\
turbo & \includegraphics[height=1.6ex,width=1.1cm]{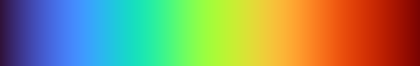} & 0.549 & 0.527 & 0.54 & 0.452 & 0.346 \\
vanimo & \includegraphics[height=1.6ex,width=1.1cm]{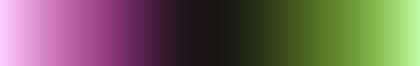} & 0.502 & 0.728 & 0.635 & 0.31 & 0.345 \\
viridis & \includegraphics[height=1.6ex,width=1.1cm]{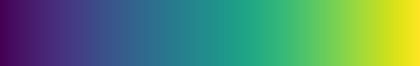} & 0.299 & 0.562 & 0.441 & 0.463 & 0.485 \\
\bottomrule
\end{tabular}
\end{table}

\end{document}